\documentstyle[psfig,epsf]{mn}

\newif\ifAMStwofonts
\AMStwofontstrue

\ifoldfss
  \ifCUPmtlplainloaded \else
    \NewTextAlphabet{textbfit} {cmbxti10} {}
    \NewTextAlphabet{textbfss} {cmssbx10} {}
    \NewMathAlphabet{mathbfit} {cmbxti10} {} 
    \NewMathAlphabet{mathbfss} {cmssbx10} {} 
  \fi
  \ifAMStwofonts
    \ifCUPmtlplainloaded \else
      \NewSymbolFont{upmath} {eurm10}
      \NewSymbolFont{AMSa} {msam10}
      \NewMathSymbol{\upi}     {0}{upmath}{19}
      \NewMathSymbol{\umu}     {0}{upmath}{16}
      \NewMathSymbol{\upartial}{0}{upmath}{40}
      \NewMathSymbol{\leqslant}{3}{AMSa}{36}
      \NewMathSymbol{\geqslant}{3}{AMSa}{3E}

      \let\leq=\leqslant \let\le=\leqslant
      \let\geq=\geqslant \let\ge=\geqslant
    \fi
  \fi
\fi 

\ifnfssone
  \newmathalphabet{\mathit}
  \addtoversion{normal}{\mathit}{cmr}{m}{it}
  \addtoversion{bold}{\mathit}{cmr}{bx}{it}
  \newmathalphabet{\mathbfit} 
  \addtoversion{normal}{\mathbfit}{cmr}{bx}{it}
  \addtoversion{bold}{\mathbfit}{cmr}{bx}{it}
  \newmathalphabet{\mathbfss} 
  \addtoversion{normal}{\mathbfss}{cmss}{bx}{n}
  \addtoversion{bold}{\mathbfss}{cmss}{bx}{n}
  \ifAMStwofonts
    \ifCUPmtlplainloaded \else
      \UseAMStwoboldmath
      \makeatletter
      \new@mathgroup\upmath@group
      \define@mathgroup\mv@normal\upmath@group{eur}{m}{n}
      \define@mathgroup\mv@bold\upmath@group{eur}{b}{n}
      \edef\UPM{\hexnumber\upmath@group}
      \new@mathgroup\amsa@group
      \define@mathgroup\mv@normal\amsa@group{msa}{m}{n}
      \define@mathgroup\mv@bold\amsa@group{msa}{m}{n}
      \edef\AMSa{\hexnumber\amsa@group}
      \makeatother
      \mathchardef\upi="0\UPM19
      \mathchardef\umu="0\UPM16
      \mathchardef\upartial="0\UPM40
      \mathchardef\leqslant="3\AMSa36
      \mathchardef\geqslant="3\AMSa3E

      \let\leq=\leqslant \let\le=\leqslant
      \let\geq=\geqslant \let\ge=\geqslant
    \fi
  \fi
\fi 

\ifnfsstwo
  \DeclareMathAlphabet{\mathbfit}{OT1}{cmr}{bx}{it}
  \SetMathAlphabet\mathbfit{bold}{OT1}{cmr}{bx}{it}
  \DeclareMathAlphabet{\mathbfss}{OT1}{cmss}{bx}{n}
  \SetMathAlphabet\mathbfss{bold}{OT1}{cmss}{bx}{n}
  \ifAMStwofonts
    \ifCUPmtlplainloaded \else
      \DeclareSymbolFont{UPM}{U}{eur}{m}{n}
      \SetSymbolFont{UPM}{bold}{U}{eur}{b}{n}
      \DeclareSymbolFont{AMSa}{U}{msa}{m}{n}
      \DeclareMathSymbol{\upi}{0}{UPM}{"19}
      \DeclareMathSymbol{\umu}{0}{UPM}{"16}
      \DeclareMathSymbol{\upartial}{0}{UPM}{"40}
      \DeclareMathSymbol{\leqslant}{3}{AMSa}{"36}
      \DeclareMathSymbol{\geqslant}{3}{AMSa}{"3E}

      \let\leq=\leqslant \let\le=\leqslant
      \let\geq=\geqslant \let\ge=\geqslant
    \fi
  \fi
\fi 

\ifCUPmtlplainloaded \else
  \ifAMStwofonts \else 
    \def\upi{\pi}
    \def\umu{\mu}
    \def\upartial{\partial}
  \fi
\fi

\title[FSVS Cluster Catalogue]{The FSVS Cluster Catalogue: Galaxy Clusters and
Groups in the Faint Sky Variability Survey}

\author[S\"{o}chting et al.]
{Ilona K. S\"{o}chting,$^{1,2}$ Mark E. Huber,$^{3,4}$ Roger G. Clowes,$^5$
Steve B. Howell$^6$ \\
$^1$ Isaac Newton Group, Apartado de Correos 321, 38700 Santa Cruz de 
La Palma, Canary Islands, Spain \\
$^2$ Astrophysics, Denys Wilkinson Building, Keble Road, Oxford OX1 3RH, UK, (E-mail: iks@astro.ox.ac.uk) \\
$^3$ Department of Physics and Astronomy, University of British Columbia,
6224 Agricultural Road, Vancouver, BC, V6T 1Z1, Canada \\
$^4$ Lawrence Livermore National Laboratory, P.O. Box 808, L-413, Livermore, CA 94551, USA \\
$^5$ Computational Astrophysics, Department of Computing, University of
Central Lancashire, Preston, PR1 2HE, UK \\
$^6$ WIYN Observatory and National Optical Astronomy Observatory
950 N. Cherry Ave., Tucson, AZ 85726, USA }

\date{Accepted 2006 Xxxxxxxx xx.
      Received 2006 Xxxxxxxx xx;
      in original form 2005 Xxxxxxxx xx}

\pagerange{\pageref{firstpage}--\pageref{lastpage}}
\pubyear{2006}

\begin{document}

\maketitle

\label{firstpage}

\begin{abstract}

We describe a large sample of 598 galaxy clusters and rich groups discovered
in the data of the Faint Sky Variability Survey. The clusters have been
identified using a fully automated, semi-parametric technique based on a
maximum likelihood approach applied to Voronoi tessellation, and enhanced by
colour discrimination. The sample covers a wide range of richness, has a
density of $\sim 28$ clusters deg$^{-2}$, and spans a range of estimated
redshifts of $0.05 < z < 0.9$ with mean $\langle z \rangle = 0.345$. Assuming
the presence of a cluster red sequence, the uncertainty of the estimated
cluster redshifts is assessed to be $\sigma \sim 0.03$. Containing over 100
clusters with $z > 0.6$, the catalogue contributes substantially to the
current total of optically-selected, intermediate-redshift clusters, and
complements the existing, usually X-ray selected, samples. The FSVS fields
are accessible for observation throughout the whole year, making them
particularly suited for large follow-up programmes. The construction of this
{\it FSVS Cluster Catalogue\/} completes a fundamental component of our
continuing programmes to investigate the environments of quasars and the
chemical evolution of galaxies. We publish here the list of all clusters with
their basic parameters, and discuss some illustrative examples in more
detail. The full {\it FSVS Cluster Catalogue,} together with images and lists
of member galaxies etc., will be issued as part of the ``NOAO data
products'', and accessible at http://www.noao.edu/dpp/. We describe the
format of these data and access to them.

\end{abstract}


\section{Introduction}

Galaxy clusters are important for the study of structure formation and
evolution, and their distribution provides important constraints on
cosmological models. The number density of rich galaxy clusters is sensitive
not only to the comoving distance measure but also to the rate of growth of
fluctuations and hence the matter density $\Omega_{m}$. The low matter
density models predict larger fluctuations at earlier times, because the
growth of structure ceases earlier. Using the Press-Schechter formalism
\cite{PS74} this predicts more clusters of a given mass at high
redshift. However, large uncertainties arise when using cluster surveys to
constrain cosmological parameters due to systematic uncertainties in the
modelling of cluster formation and evolution \cite{VL99}. Any progress in
this field depends strongly on the availability of large cluster samples
covering a wide parameter space, which require a multiplicity of selection
methods. This justifies the large effort being invested into detection of
these largest gravitationally bound systems by probing different regions of
the electromagnetic spectrum.

Following the analysis of the first-year WMAP data the cosmological model can
be studied with very high detail (Spergel et al.\ 2003) and combining the
WMAP data with that of large surveys of galaxies (e.g. 2dF and SDSS), the
cosmological parameters can be constrained with high precision (Hawkins et
al.\ 2003, Percival et al. 2002). Nevertheless, the cosmological model
remains undecided in the sense that the cause of the accelerated expansion is
unknown. Parker, Komp \& Vanzella (2003), for example, point out the power of
counts of clusters (or galaxies) as a function of redshift to address this
fundamental issue.

Abell (1958) constructed the first cluster catalogue by a systematic approach
to the visual inspection of photographic plates. Zwicky et al.\ (1961--1968)
constructed another large catalogue, also using visual inspection.
Improvement of the performance and accessibility of computers allowed the
implementation of fully automated cluster algorithms (e.g.\ Shectman 1985,
Dodd \& MacGillivray 1986). Since spectroscopic information is limited to
very small areas of sky or to low redshifts (e.g $z \sim 0.15$ for the
2dFGRS, Colless et al.\ 2001), the challenge for cluster detection algorithms
is to reduce the projection effects using only photometric data. Postman et
al.\ (1996) introduced a matched filter (MF) algorithm --- a
maximum-likelihood method, which assumes a filter for both the cluster radial
profile and the luminosity function of the cluster galaxies. At the same time
as improvements to the original MF resulted in the adaptive matched filter
(AMF, Kepner et al.\ 1999), many other statistical and astronomical concepts
found applications in galaxy cluster surveys. Voronoi tessellation (VT) has
been applied very successfully in connection with thresholding of the density
peaks (Ramella et al.\ 2001, S\"{o}chting, Clowes \& Campusano 2002, Kim et
al.\ 2000), and recently improved by incorporating a maximum likelihood
estimator (MLE, S\"{o}chting, Clowes \& Campusano 2004) and colour
discrimination (Kim et al.\ 2002). Colour discrimination is based on the fact
that members of all known galaxy clusters trace a narrow sequence in
colour-magnitude --- the cluster red sequence (CRS) ---, which was first
proposed by Gladders \& Yee (2000) as a means to increase the contrast of
galaxy clusters above background galaxies.

We have used the images and object catalogues from the Faint Sky Variability
Survey (FSVS, Groot et al.\ 2003) to identify galaxy clusters and groups and
hence create the {\it FSVS Cluster Catalogue.} (Note that our application,
being very different from that for which the FSVS was created, is one example
of the usefulness of the ``Virtual Observatory.'') The primary motivations
for finding galaxy clusters in the FSVS data are the study of: (i) the galaxy
environments favouring the formation of quasars, and hence the main
mechanisms of formation as a function of redshift; and (ii) the chemical
evolution of galaxies and dependences on environment. We chose VT, enhanced
by MLE and a colour-cut, to construct a morphologically-unbiased sample
(Okabe et al.\ 2000, Allard \& Fraley 1997), containing structures with a
wide range of richnesses and evolutionary stages. The resulting cluster
catalogue, containing $\sim 600$ groups and clusters at $0.05 < z < 0.9$,
complements the existing samples which at intermediate redshifts are usually
biased towards X-ray luminous and/or optically very rich clusters.

We describe the data from the FSVS in Section 2 and the cluster detection
technique in Section 3. In Section 4 we present the properties of the {\it
FSVS Cluster Catalogue.} Section 5 contains the basic tabular data for all
FSVS clusters and describes briefly the on-line access to the complete
catalogue data.

\section{Data}

\begin{figure}
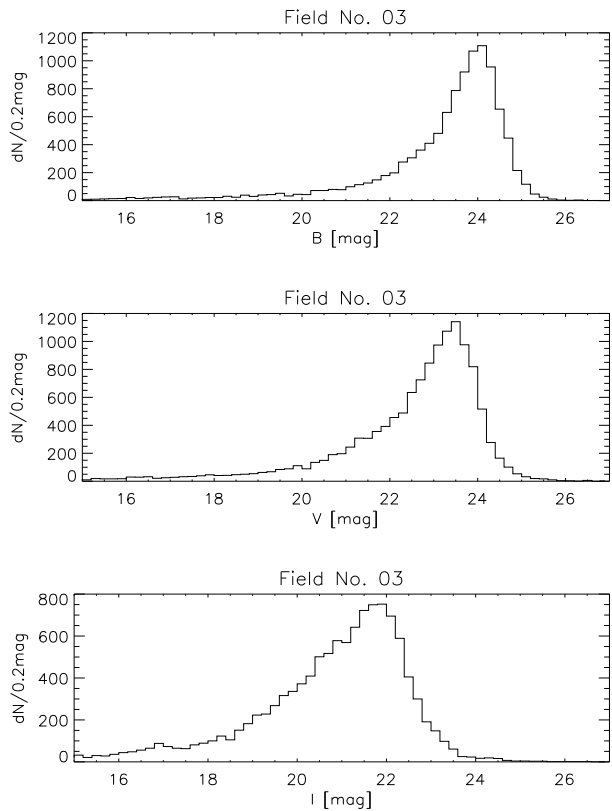

\psfig{file=figure1a.epsi,width=8.0cm}
\vspace{0.5cm} \psfig{file=figure1b.epsi,width=8.0cm}
\vspace{0.5cm} \psfig{file=figure1c.epsi,width=8.0cm}
\caption{The magnitude distribution of the FSVS data in $B$ (top), $V$
(middle) and $I$ (bottom) using Field 3 as an illustrative example. All
objects, point-like and extended, have been included.}
\label{mag_distr}
\end{figure}

The FSVS provides a unique combination of spatial coverage and
depth in three optical passbands with very high photometric and astrometric
accuracy. The survey used the 2.5 metre Isaac Newton Telescope (INT) on La
Palma to cover, in $B$, $V$ and $I$, 79 fields of moderate to high galactic
latitude sky (see Table~\ref{fsvs_fields}), totalling $\sim 23$ deg$^{2}$.
The limiting magnitudes are $B \sim 24.5$ mag., $V \sim 25$ mag., and $I \sim
24$ mag. The completeness reaches a peak at $B \sim 24.0$ mag., $V \sim 23.5$
mag., and $I \sim 21.5$ mag. (Figure~\ref{mag_distr}). Owing to poor quality,
four fields (nos. 7, 36, 51, 67) have been excluded from this investigation,
leading to a final coverage of $21.75$ deg$^{2}$. For more information on the
FSVS data products see Groot et al.\ (2003) and Huber (2002). The photometry
database for the FSVS is currently available for on-line access at {\it
http://www.astro.uva.nl/$\sim$fsvs/}.

\begin{table}
\center
\caption{The approximate positions of the FSVS fields (RA, Dec and
$l^{II}$, $b^{II}$) and contiguous areas occupied by field
groupings. (From Huber 2002.)}
\label{fsvs_fields}
\begin{tabular}{llllll}
\hline
Field No.  & RA      & Dec      & $l^{II}$ & $b^{II}$ & area        \\
           & (J2000) & (J2000)  & [deg]    & [deg]    & [deg$^{2}$] \\
\hline
1--6       & 23:44   & $+$27:15 & 105      & $-$33    & 1.74        \\
8--12      & 02:32   & $+$15:00 & 156      & $-$40    & 1.45        \\
13--18     & 07:52   & $+$20:40 & 200      & $+$22    & 1.74        \\
19--22     & 12:53   & $+$27:01 & 220      & $+$90    & 1.16        \\
23--26     & 12:51   & $+$26:20 & 268--360 & $+$89    & 1.16        \\
27--30     & 16:25   & $+$26:33 & 45       & $+$43    & 1.16        \\
31--34     & 17:20   & $+$27:00 & 49       & $+$31    & 1.16        \\
35, 37--40 & 03:02   & $+$18:38 & 161      & $-$33    & 1.45        \\
41--46, 59 & 07:15   & $+$21:00 & 196      & $+$15    & 2.03        \\
47--50, 60 & 10:00   & $+$21:30 & 211      & $+$50    & 1.45        \\
52--56     & 16:23   & $+$27:03 & 45       & $+$42    & 1.45        \\
57--58     & 16:32   & $+$21:16 & 39       & $+$39    & 0.58        \\
61--62     & 10:37   & $+$04:00 & 242      & $+$50    & 0.58        \\
63--66     & 17:25   & $+$27:30 & 50       & $+$30    & 1.16        \\
68--71     & 22:02   & $+$27:30 & 83       & $-$21    & 1.16        \\
72--75     & 18:32   & $+$36:00 & 64       & $+$19    & 1.16        \\
76--79     & 23:47   & $+$28:10 & 106      & $-$32    & 1.16        \\
\hline
\end{tabular}
\end{table}

\subsection{Star-Galaxy Separation}

\begin{figure}
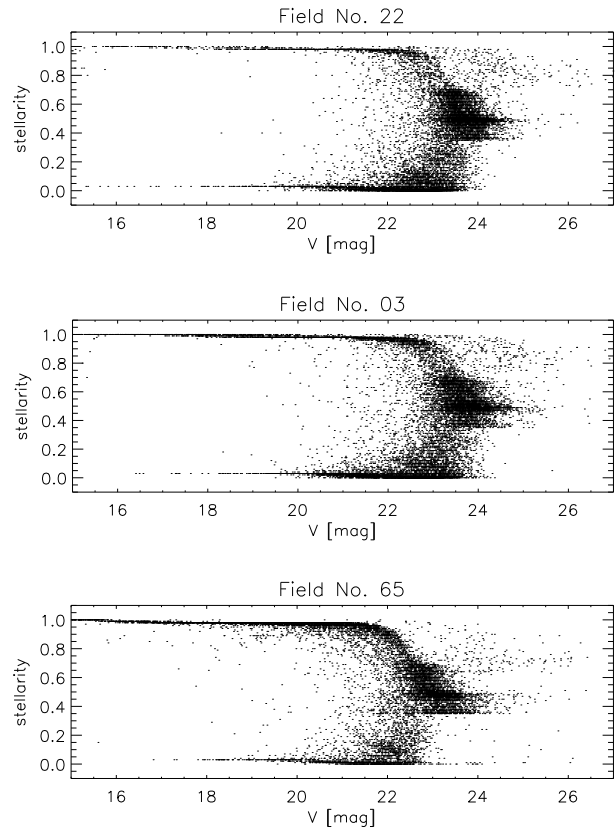

\psfig{file=figure2a.epsi,width=8.0cm}
\vspace{0.5cm} \psfig{file=figure2b.epsi,width=8.0cm}
\vspace{0.5cm} \psfig{file=figure2c.epsi,width=8.0cm}
\caption{The stellarity as a function of magnitude in $V$. To illustrate the
variation of the quality of co-added $V$ frames, the best frame (Field 22),
(in terms of the depth, resolution and galactic latitude), average frame
(Field 03), and worst frame (Field 65) have been shown.}
\label{fsvs_stellarity}
\end{figure}

Star-galaxy separation is carried out using the parameter {\it stellarity},
from the source detection program SExtractor \cite{BA96}. It is based on the
extended nature of the identified source and can have values ranging from 0
to 1, with 1 corresponding to a point-source. In Figure~\ref{fsvs_stellarity}
we have plotted the stellarity as a function of magnitude for $V$ data,
which, benefiting from the variability sampling, has improved the
signal-to-noise ratio using co-added multiple field pointings. The quality
does vary from field to field, which is best illustrated by comparison of the
best (Field 22), in terms of the depth, resolution and galactic latitude, and
worst (Field 65) fields in the top and bottom plots of
Figure~\ref{fsvs_stellarity}.

\begin{figure}
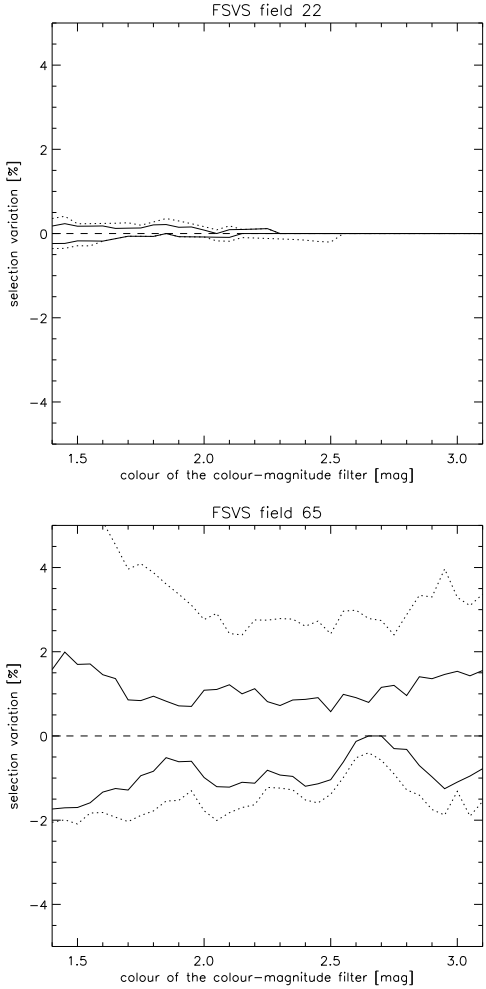

\psfig{file=figure3a.epsi,width=7.0cm}
\psfig{file=figure3b.epsi,width=7.0cm}
\caption{The impact of the threshold stellarity on the object selection. The
increment of galaxy-like objects selected for stellarity values of 0.70,
0.75, 0.8, 0.85 and 0.90 (upper dotted, upper solid, dashed, lower solid and
lower dotted lines respectively) is expressed in percent relative to the
number of objects selected for stellarity value of 0.8. The effect is plotted
as a function of the colour of the colour-magnitude filter (the CM filter
will be described in detail later in this paper). To illustrate the variation
of the quality of co-added $V$ frames, the best frame (top, Field 22) and
worst one (bottom, Field 65) have been shown.}
\label{stel_distr_norm}
\end{figure}

There is no obvious cutoff between point and extended sources but the
empirical data from follow-up spectroscopy and matching to the SIMBAD/VizieR
database supports the threshold of 0.8. It provides the best separation
(Huber 2002) between stars ({\it stellarity} $\geq 0.8$) and galaxies ({\it
stellarity} $<0.8$) to a high magnitude ($V \sim 22.0$). As shown in
Figure~\ref{stel_distr_norm}, within the range of 0.75 to 0.85, the exact
value of the threshold stellarity has little impact on the object
selection. The number of selected galaxy-like objects varies by $<1\%$ in
most of the fields and $<3\%$ in fields with data of lowest quality. Beyond
these limits, however, the number of selected galaxy-like objects start to
change more rapidly (depending also on the data quality for a given field)
and may influence the contrast of galaxy structures above the background and
consequently their detection rate. The redder objects (which are important
for our study) appear to be much less sensitive to the choice of the
stellarity threshold than bluer objects.

\subsection{Stellar Contamination}

\begin{figure}
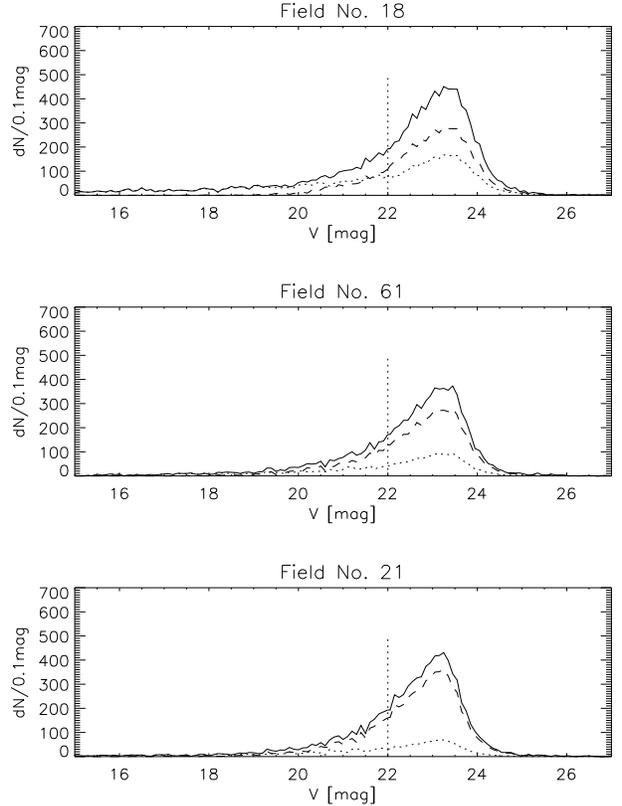

\psfig{file=figure4a.epsi,width=8.0cm}
\vspace{0.5cm} \psfig{file=figure4b.epsi,width=8.0cm}
\vspace{0.5cm} \psfig{file=figure4c.epsi,width=8.0cm}
\caption{The distributions with magnitude of the point and extended sources,
and their extrapolated values for $V > 22.0$, using Fields 18, 61 and 21 as
illustrative examples. These fields have, respectively, $b^{II} =
+22^{\circ}$, $b^{II} = +50^{\circ}$ and $b^{II} = +90^{\circ}$. The stellar
contamination is decreasing with increasing galactic latitude. The
distribution of all objects is denoted by the solid line, extended sources by
the dashed line, and point-like sources by the dotted line. The vertical
dotted line indicates the $V=22.0$ transition point to extrapolated values.}
\label{fsvs_stars}
\end{figure}

At the faint end of the data ($V > 22.0$), the star-galaxy separation is
increasingly unreliable and more and more stars are classified as extended
objects. The relative contamination by stars at $V > 22.0$ among faint
extended sources is expected to vary as a function of the galactic latitude
owing to variation of the typical star counts. It can also vary from field to
field owing to enhanced numbers of galaxies where rich clusters occur. For
these reasons the contamination by stars at $V > 22.0$ has been estimated for
every field separately. Using the magnitude distribution of point and
extended sources at $V \leq 22.0$ the relative contribution of point and
extended sources has been determined as a function of magnitude.
Figure~\ref{fsvs_stars} gives the magnitude distributions of all sources,
point sources, and extended sources using Fields 18, 61 and 21, with galactic
latitudes $+22^{\circ}$, $+50^{\circ}$, and $+90^{\circ}$, as illustrative
examples. The distributions at $22.0 < V < 25.0$ are extrapolated from
exponential fits of the point-source and extended distributions for $V \le
22.0$, then scaled, for $V > 22.0$, according to the overall numbers of
detected objects. The distributions show an obvious dependence of the stellar
contamination on the galactic latitude. However, even at the lowest
latitudes, galaxies substantially outnumber stars at the faint limits of the
data ($V > 22.0$). Under the assumption that the star population at such
faint magnitudes will comprise only a minority of the objects and that their
spatial distributions are fairly uniform compared with those of the galaxies
(see Jones et al.\ 1991 for the variance of star and galaxy counts), all
faint objects satisfying {\it stellarity} $< 0.8$ have been included in the
master database for identifying galaxy clusters.

\section{Method}

Our main concern in the selection of clusters was the avoidance, as far as
possible, of selection bias by using non-parametric methods to minimise the
assumptions about the properties of the clusters. S\"{o}chting et al.\ (2004)
proposed Voronoi tessellation (VT) enhanced by a maximum likelihood estimator
(MLE) to better delineate the boundaries of the clusters, plus colour
discrimination to reduce the contamination from background galaxies. The use
of colour information provides the additional bonus of relatively accurate
cluster redshifts deduced from the cluster red sequence (CRS, Gladders \& Yee
2000). 

The technique has been described in detail by S\"{o}chting et al.\ (2004),
but with application to SuperCosmos $B_{J} - R$ data, which, of course, is
intrinsically different from FSVS data. The SuperCosmos data cover the whole
of the southern sky using shallow ($B_{J} < 23$ and $R < 21.5$) photographic
plates whereas the FSVS data cover a relatively small area with deep ($B <
24.5$, $V <25$ and $I<24$) CCD exposures. The $B_{J},~R$ SuperCosmos data
allow us to probe the redshift range $z < 0.3$ across wide areas of sky,
whereas the FSVS Cluster Catalogue can reach $z\sim 1$ but only across
relatively small fields. Consequently, the low redshift ($z<0.3$) galaxy
clusters and superclusters are best detected using the SuperCosmos Sky Survey
(or, for example, the Sloan Digital Sky Survey in the north) whereas the FSVS
data are best suited to finding galaxy clusters at intermediate redshift
($0.3<z<1$) and poor groups at $z<0.3$.

These differences demand enhancements beyond the S\"{o}chting et al.\ (2004)
method, focusing on using colour discrimination to improve the contrast
enhancement over the large range of redshifts covered by the FSVS data. The
Monte Carlo test of the detection rate (including completness and false
detections) published in S\"{o}chting et al.\ (2004) as a function of
contrast are independent of colour and remain fully valid for the FSVS data.

\subsection{Voronoi Tessellation Technique}

\begin{figure}
\psfig{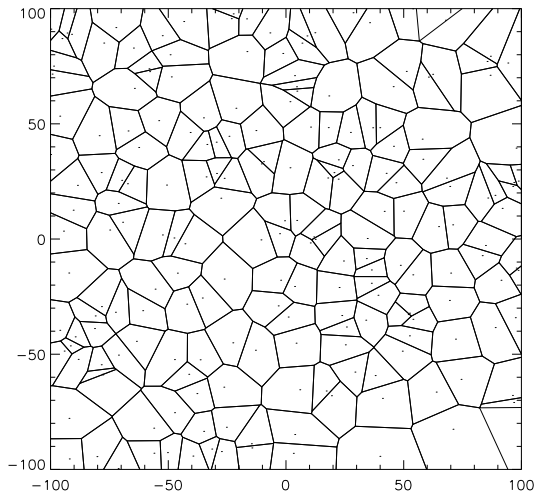}
\caption{Example of Voronoi tessellation constructed on a uniformly
distributed point set (from S\"{o}chting et al.\ 2003). For more information
on Voronoi tessellation see Okabe et al.\ (2000) and references therein. }
\label{voronoi}
\end{figure}

VT provides a partition of the investigated area into convex cells around
every galaxy (Figure~\ref{voronoi}). The inverse of the area of a Voronoi
cell gives the number density at the position of the galaxy. Since only the
spatial structure of the galaxy distribution decides the sizes of the cells,
VT provides a non-parametric method of sampling the underlying density
distribution.

Galaxy clusters are detected as peaks in the galaxy density ($\delta$)
distribution. The simplest approach to locate the density peaks is to select
objects that exceed a threshold $\sigma$ for the density contrast with
respect to the background. The density contrast $\sigma_{i}$ at the position
of the $i$th object is defined as
\begin{equation}
\sigma_{i} = (\delta_{i} - \bar{\delta}) / \bar{\delta}, 
\end{equation}
where $\delta_{i}$ is the density and $\bar{\delta}$ is the mean density.
One should remember that using Voronoi cells the mean density is calculated as
\begin{equation}
\bar{\delta} = \frac{1}{n} \sum_{i=1}^{n} \frac{1}{A_{i}}
\end{equation}
where $A_{i}$ is the area of the Voronoi cell around object $i$ and $n$ is
the overall number of objects. Until now this approach has been applied in
all VT-based procedures, producing good results
\cite{Ra01,Ki02,So02}. Thresholding, however, introduces a bias towards
clusters in regions of enhanced background, because it is not locally
adaptive. S\"{o}chting et al.\ (2004) proposed an enhancement to this
approach through a maximum likelihood estimator (MLE). Following the
mathematical framework proposed by Allard \& Fraley (1997), the
log-partial-maximised-likelihood of a set of galaxies to be a cluster can be
expressed as
\begin{equation}
l(\bmath{x}; {\mathcal{A}}) = -n \ln{n} + N_{A} \ln{\frac{N_{A}}{|A|}}
+ (n - N_{A})\ln{\frac{n - N_{A}}{1 - |A|}}.
\end{equation}
where $|A|$ denotes the normalised area of the cluster (the physical area of
the cluster $\mathcal{A}$ divided by the physical area of a FSVS field $K$)
and $N_{A}$ is the number of cluster member galaxies. By constructing
$\mathcal{A}$ from Voronoi cells we ensure that any spatial constraints are
defined by the data points themselves.

Without advance knowledge of the number and approximate positions of clusters
in the data set, the MLE is very computationally intensive, losing the speed
advantage of the basic VT. To accelerate the computation, the thresholding is
preserved, but as a preliminary step (to produce a candidate list for the MLE
algorithm), and MLE is used in the main process to reduce false detections
(spurious clusters) and find the member galaxies in the confirmed
clusters. The application of the MLE allows us to choose a rather low
threshold, ensuring that the poor clusters in the regions of lower background
density will be included. If the density of the cluster galaxies is
$\delta_{cl}$ and that of the background galaxies $\bar{\delta_{b}}$ then the
contrast is
\begin{equation}
\sigma = \frac{(\delta_{cl} + \bar{\delta_{b}}) - \bar{\delta_{b}}}{\bar{\delta_{b}}} = \frac{\delta_{cl}}{\bar{\delta_{b}}} .
\end{equation}
Since $\delta_{cl} = N_{A}/\mathcal{A}$ and $\bar{\delta_{b}} = (n -
N_{A})/K$ this becomes
\begin{equation}
\sigma = \frac{N_{A}}{\mathcal{A}} \frac{K}{n-N_{A}}.
\end{equation}
Then, since ${\mathcal{A}}/K = |A|$, the contrast of a cluster is
\begin{equation}
\sigma = \frac{1}{n-N_{A}} \frac{N_{A}}{|A|}
\end{equation}
Assuming that the overall number of objects in the sample is very high
compared with the number of cluster members, the size of an average Voronoi
cell in a cluster is approximately
\begin{equation}
\langle |A_{cl}| \rangle \approx \frac{1}{n \sigma}
\end{equation}
Using synthetic clusters, S\"{o}chting et al.\ (2004) have determined that the
best performance is achieved by a threshold of $\sigma = 2.0$ (i.e.\ all cells
satisfying $A_{i} \le 1/(2.0~n)$ are selected), achieving a detection rate of
the synthetic clusters close to $100\%$ with contamination by spurious
clusters less than $30\%$ of the overall number of clusters detected.

To suppress chance associations, groups with fewer than seven members present
in the final cluster sample (after applying MLE) have to be discarded. The
minimum number of members has been dictated by basic statistical properties
of Voronoi tessellation. Assuming a Poisson distribution, the mean number of
vertices/edges of a typical cell is 6 (Okabe et al.\ 2000), marking the
natural threshold for random associations. Poor galaxy groups, which are all
affected by this constraint, will be addressed in future work, using a
somewhat different technique.

\subsection{Colour Discrimination}

\begin{figure}
\psfig{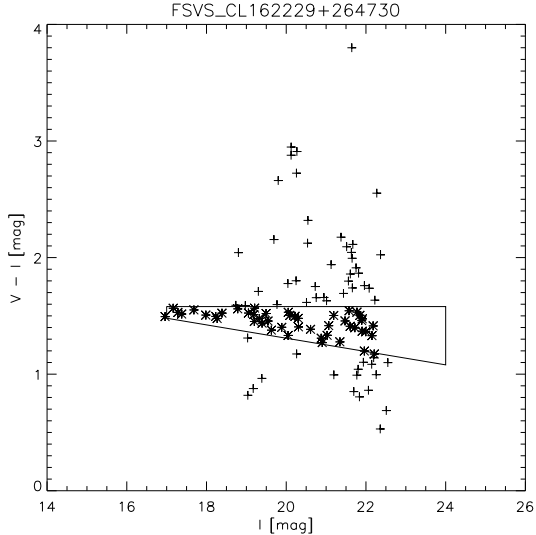}
\caption{Example of a colour-magnitude filter (the polygon) used to enhance
the contrast of galaxy clusters in the FSVS data. Asterisks denote galaxies
that are identified as cluster members --- they are within both the cluster
boundary and the colour-magnitude filter; pluses denote galaxies within the
cluster boundary but outside the colour-magnitude filter.}
\label{new_filter}
\end{figure}

The cluster detection rate and the contamination by spurious clusters depend
strongly on the contrast of clusters above background, which can be increased
by applying a colour discrimination. The use of colour slices was proposed
and tested by Gladders \& Yee (2000). The colour-magnitude slices used by
S\"{o}chting et al.\ (2004) with $B_{J}$ and $R$ data have proved appropriate
for detecting galaxy clusters at low redshifts ($z < 0.3$), where the slopes
and zero-points of the CRS are well known. At higher redshifts, the majority
of the known galaxy clusters have been discovered by X-rays, and only a few
of them have well-studied photometric properties. Beyond $z \sim 0.5$ the
theoretically predicted parameters of the CRS are highly dependent on the
selected models \cite{Ko98}, and would introduce a strong bias if
incorporated in an algorithm for detecting clusters. The wide range of
redshifts expected to be covered with FSVS data requires the introduction of
a broader colour-magnitude filter to accommodate the strong variation of the
parameters of the CRS. The shape of the filter must accommodate: (1) at low
redshift ($z < 0.3$) the population of faint cluster galaxies, for which the
colour distribution is broader than that of the bright cluster spheroids; (2)
at higher redshift ($z > 0.3$), where only the bright cluster galaxies are
detected, the uncertainty in the slope of the CRS. Figure~\ref{new_filter}
illustrates the shape of the filter we have adopted.

The filter is narrow at the bright end of the $I$ axis and gradually widens
towards bluer colours at the faint end, enclosing a locus known from the
galaxy distribution in low $z$ clusters. The purpose of the filter is to
maximize, at any given redshift, the contrast of a galaxy cluster above the
background, sometimes at the expense of some cluster galaxies staying outside
the selection (e.g.\ the brightest members of the richest clusters at $z <
0.2$, or a large fraction of the blue late-type galaxies). By moving the
filter up the colour axis in $\Delta (V - I) = 0.05$ steps, higher redshifts
are probed. The filter is not moved along the luminosity axis ($I$), contrary
to any expected allowance for galaxies becoming fainter with increasing
redshift. Instead, the position of the filter remains unchanged with respect
to the $I$ axis, allowing the bright galaxies of the higher redshift clusters
to fall within the wider region of the filter, which accommodates a wide
range of CRS slopes.

\subsection{The Cluster Detection Algorithm}

The MLE procedure to find galaxy clusters has been applied to every FSVS
field separately. No attempts have been made to apply it across the
boundaries of adjacent fields because of potential difficulties if the depths
are different. In most cases, clusters that cross boundaries will be
recognised and united in the later stages when multiple detections are
consolidated.

The detection algorithm consists of the following steps. (1) Select all
objects for which ($V-I$) and $I$ satisfy the criteria of the first colour
filter. (2) Calculate the Voronoi tessellation for this selection. (3) Find
which Voronoi cells satisfy $A_{i} \le 1/(2.0~n)$. These are the cluster
candidates. Multiple adjacent cells satisfying the criterion are treated as
one candidate. (4) Apply the MLE to every candidate cluster. If the candidate
consists of multiple cells then start with the smallest cell. (5) Move the
colour filter by $\Delta (V-I) = 0.05$ and repeat steps 1--4 until the whole
colour space is covered. (6) Combine clusters that share the majority of the
members. (7) For every cluster, place the colour filter to cover the CRS
marked by its members. Select all objects within this filter and the cluster
boundary (defined as the smallest convex hole enclosing all Voronoi tessells
of the members) as the final cluster members. (8) Discard clusters with fewer
than seven members.

The above procedure is sensitive to clusters that are truncated by the field
boundary if at least seven cluster galaxies have Voronoi cells with edges
that do not intersect the field boundary. In contrast, those methods that
match a pre-defined radial profile are ineffective. Clusters detected at the
field boundaries are flagged in the on-line catalogue, since some of their
parameters (e.g.\ richness and radius) reflect only the fraction of the
cluster contained by the FSVS field.

\subsection{Calibration of Cluster Redshifts}

The redshifts of the clusters can be estimated empirically, using the colour
of the CRS, if some of the clusters have known spectroscopic redshifts.
Ideally, the spectroscopic redshifts should cover the whole redshift range
accessible by the survey. However, only four galaxy clusters with published
spectroscopic redshifts were originally covered by the FSVS data, which would
provide only a restricted calibration sequence. To construct a continuous
sequence, redshifts of five additional clusters were obtained through the
service programme at the Wiliam Herschel Telescope using the ISIS
instrument. Data reduction was performed using standard IRAF routines. The
combined sample used for redshift calibration is listed in
Table~\ref{calib_clst}.

\begin{table}
\center
\caption[FSVS clusters with spectroscopic redshifts]{\footnotesize{FSVS
clusters with spectroscopic redshifts used for the redshift calibration of
the cluster colours. The spectroscopic redshifts originate from the following
sources: (1) Mullis et al.\ 2003; (2) Popesso et al.\ 2004; (3) Novicki,
Sornig \& Henry 2002; and (4) WHT ISIS service observations.}}
\label{calib_clst}
\begin{tabular}{llll}
\hline
Cluster ID                & $z$   & $C_{cl}$ [$V-I$ mag] & Ref. \\
\hline
FSVS$\_$CL163105$+$212141 & 0.098 & 1.48                 & 1    \\
FSVS$\_$CL172015$+$263858 & 0.161 & 1.40                 & 2    \\
FSVS$\_$CL172030$+$274013 & 0.164 & 1.62                 & 2    \\
FSVS$\_$CL023057$+$144500 & 0.36  & 2.11                 & 4    \\
FSVS$\_$CL162339$+$263433 & 0.427 & 2.13                 & 3    \\
FSVS$\_$CL220532$+$270715 & 0.50  & 2.35                 & 4    \\ 
FSVS$\_$CL172531$+$274247 & 0.56  & 2.56                 & 4    \\ 
FSVS$\_$CL071535$+$212351 & 0.81  & 2.89                 & 4    \\ 
FSVS$\_$CL234150$+$265516 & 0.93  & 3.15                 & 4    \\
\hline
\end{tabular}
\end{table}

The second challenge is to find appropriate point of reference within the colour distribution of the cluster members which would allow robust colour-to-redshift association independent of cluster richness. At low redshift, often the zero-point of the CRS or the colour at a fixed magnitude along the CRS was selected as such a point of reference. However, the slopes of the CRS
vary as a function of redshift and also as a function of richness
\cite{So02a}, plus the uncertainty in the measurement of the slope
increases with redshift because of the limited range of available
magnitudes. For these reasons we use the colour $C_{fl}$ of the
colour-magnitude filter (defined as the maximum colour covered by the filter)
in which a cluster has been detected instead of the zero-point of the
CRS. For the majority of the rich clusters both values agree very closely.

\begin{figure}
\psfig{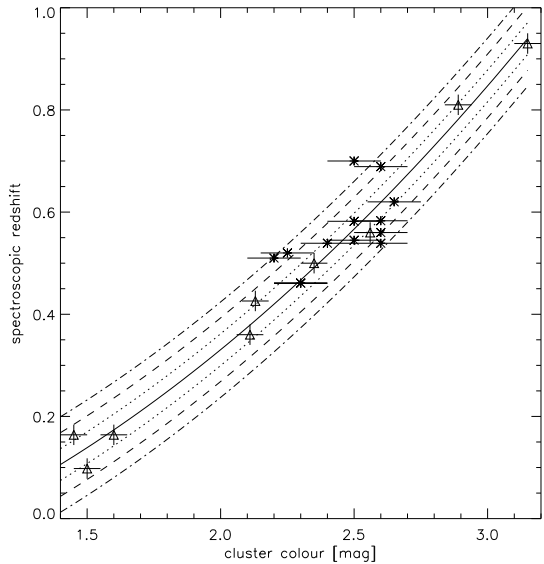}
\caption{The redshift calibration curve. ``Cluster colour'' refers to the
colour of the colour-magnitude filter (defined by its horizontal upper limit)
in which the cluster has been detected. It is also the maximum colour allowed
for a member galaxy. Triangles indicate nine calibration points (FSVS clusters
with spectroscopic redshifts) including corresponding error bars. To provide
an independent test of the calibration sequence, 13 galaxy clusters with
published $VI$ magnitudes (Stanford et al.\ 2002) and spectroscopic redshifts
have been included (asterisks). Dotted, dashed and dot-dashed lines indicate
$1\sigma$, $2\sigma$ and $3\sigma$ limits respectively.}
\label{calib_fsvs}
\end{figure}

The resulting calibration function for a polynomial fit to spectroscopic
redshifts as a function of colour for 9 FSVS clusters
(Figure~\ref{calib_fsvs}) is
\begin{equation}
z_{est} = -0.171 + 0.074C_{fl} + 0.088C_{fl}^{2}
\end{equation}

The redshift uncertainty, expressed as the standard deviation of the
residuals of the calibration points, is $\sigma = 0.030$. The uncertainty of
the cluster colour ($\Delta C_{fl}=0.05$) translates to a contribution of
0.004 to 0.013 in the redshift range $0.0 < z < 1.0$. To provide an
independent test of the redshift calibration, a sample of intermediate
redshift clusters with published spectroscopic redshifts and $VI$ photometric
data has been selected from the Stanford et al.\ (2002) catalogue
(Table~\ref{stanford}). The distribution of the reference clusters
(Figure~\ref{calib_fsvs}) shows a redshift offset of 0.014 and a standard
deviation of $\sigma=0.063$. However, the uncertainty of the colour term for
the reference clusters is very high ($\Delta C_{fl}~ \sim 0.1$) relative to
the FSVS clusters, which benefit from millimagnitude photometric accuracy
(one of the goals of the FSVS was to study microvariability).

\begin{table}
\center
\caption{Sample of reference clusters selected from the Stanford et al.\
(2002) catalogue to provide an independent test of the redshift calibration.}
\label{stanford}
\begin{tabular}{lll}
\hline
Cluster ID       & $z$   & $C_{cl}$ [$V-I$ mag] \\
\hline
3C 295           & 0.461 & 2.3                  \\
3C 313           & 0.461 & 2.3                  \\
F1557.19TC       & 0.510 & 2.2                  \\
Vidal 14         & 0.520 & 2.25                 \\
GHO 1601$+$4253  & 0.539 & 2.6                  \\
MS 0451.6$-$0306 & 0.539 & 2.4                  \\
Cl 0016$+$16     & 0.545 & 2.5                  \\
J1888.16CL       & 0.560 & 2.6                  \\
MS 2053.7$-$0449 & 0.582 & 2.5                  \\
GHO 0317$+$1521  & 0.583 & 2.6                  \\
3C 220.1         & 0.620 & 2.65                 \\
3C 34            & 0.689 & 2.6                  \\
GHO 2155$+$0321  & 0.700 & 2.5                  \\
\hline
\end{tabular}
\end{table}

\section{Cluster Sample}

\begin{figure}
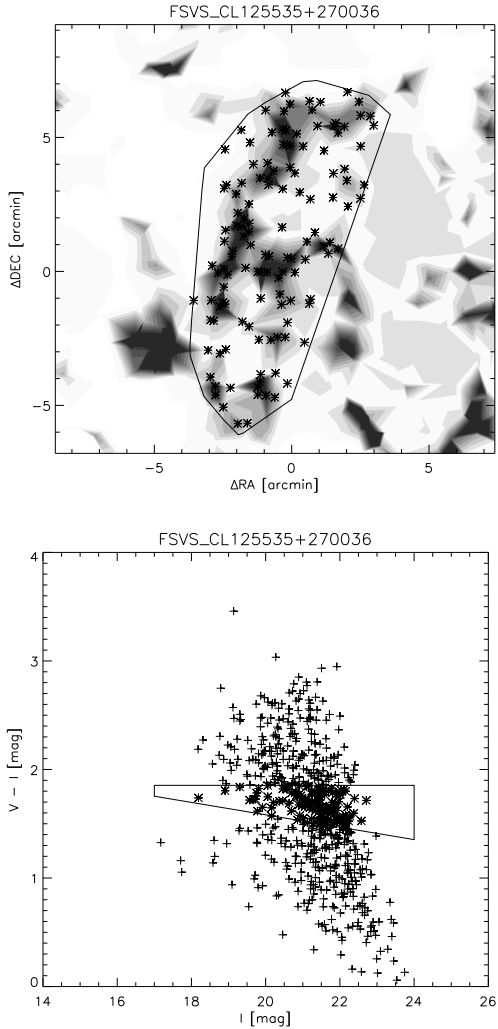

\psfig{file=figure8a.epsi,width=6.5cm}
\vspace{0.5cm}
\psfig{file=figure8b.epsi,width=6.5cm}
\caption{FSVS$\_$CL125535$+$270036 --- a highly elongated, rich (richness $=
49$; see the text for definition) galaxy cluster at $z_{est} = 0.279$,
detected in Field 22. This cluster has very prominent substructure with
galaxy groups tracing a filament-like structure. The cluster is $\sim$
$2.5h^{-1}$ Mpc by $1h^{-1}$ Mpc. Contours indicate the density of galaxies
within the colour-magnitude filter of the cluster. The black polygon
delineates the boundary of the cluster and asterisks denote its members. In
the colour-magnitude diagram: asterisks denote galaxies that are identified
as cluster members --- they are within both the cluster boundary and the
colour-magnitude filter (polygon); pluses denote galaxies within the cluster
boundary but outside the colour-magnitude filter.}
\label{clust_morph1}
\end{figure}

\begin{figure}
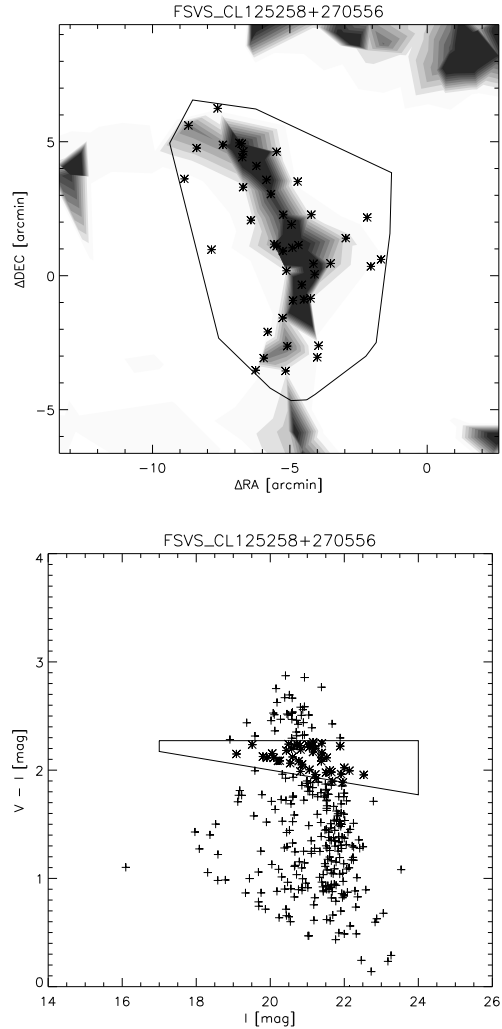

\psfig{file=figure9a.epsi,width=6.5cm}
\vspace{0.5cm}
\psfig{file=figure9b.epsi,width=6.5cm}
\caption{FSVS$\_$CL125258$+$270556 --- a highly elongated, rich (richness $=
37$) galaxy cluster at $z_{est} = 0.470$, detected in Field 21. The cluster
is $\sim$ $2.8h^{-1}$ Mpc by $1.5h^{-1}$ Mpc, but without the clumpiness of
FSVS$\_$CL125535$+$270036. Although at higher redshift ($z_{est} = 0.470$)
its third brightest galaxy is only marginally fainter than in
FSVS$\_$CL125535$+$270036, indicating higher intrinsic brightness. See the
caption to Figure~\ref{clust_morph1} for further details.}
\label{clust_morph2}
\end{figure}

\begin{figure}
\psfig{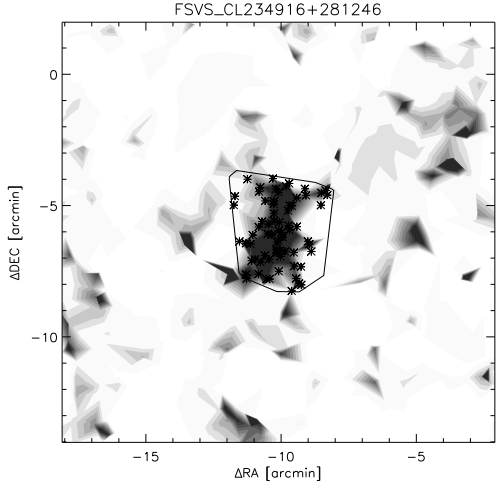}
\vspace{0.5cm}
\psfig{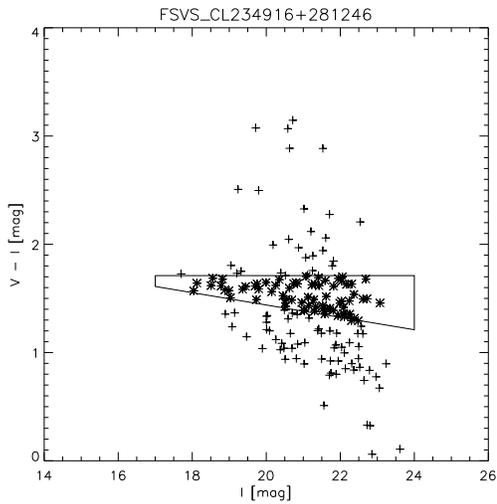}
\caption{FSVS$\_$CL234916$+$281246 --- an only marginally elongated, rich but
compact cluster at $z_{est} = 0.219$, detected in Field 79. Within the
cluster boundary, multiple compact groups are still resolved indicating their
relatively recent infall into the cluster. The cluster is $\sim$ $0.8h^{-1}$
Mpc in diameter. See the caption to Figure~\ref{clust_morph1} for further
details.}
\label{clust_morph3}
\end{figure}

\begin{figure}
\psfig{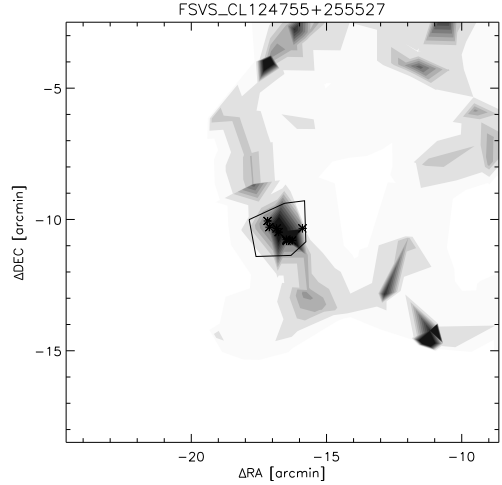}
\vspace{0.5cm}
\psfig{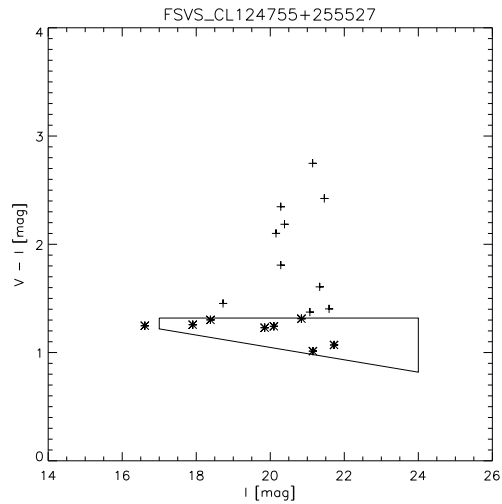}
\caption{FSVS$\_$CL124755$+$255527 --- a compact group at $z_{est} = 0.067$
with a diameter of only $\sim$ $0.1h^{-1}$ Mpc, detected in Field 26. At such
low redshifts ($z < 0.1$), the FSVS cluster catalogue does not contain very
rich and/or extended clusters owing to the relatively small size of the
fields and the method not allowing detection of galaxy clusters larger than
the window described by the data. However, considerable numbers of poor and
compact groups have been detected. Like FSVS$\_$CL124755$+$255527, many of
these groups trace a very narrow CRS in the colour-magnitude diagram,
indicating that they contain only galaxies of similar age tracing the
metallicity sequence. Such structures are very interesting subjects for
studies of the chemical evolution because clusters are assembled from
them. See the caption to Figure~\ref{clust_morph1} for further details. Note,
however, that in the colour-magnitude diagram there is an ``anomalous''
asterisk outside the colour-magnitude filter. We have added it because,
exceptionally, the brightest galaxy of the group is brighter than the bright
limit of the filter.}
\label{clust_morph4}
\end{figure}

The FSVS cluster sample consists of 598 galaxy clusters and rich groups with
$z < 1$. The cluster detection method is morphologically unbiased, resulting
in a sample of clusters with a wide range of spatial
profiles. Figures~\ref{clust_morph1} --- \ref{clust_morph4} give some
illustrative examples of the morphologies found in the sample --- radially
symmetric or elongated, compact or extended. The corresponding
colour-magnitude diagrams serve to give an impression of the richness and
brightness of these clusters.

\subsection{Richness distribution}

For our purposes, we define the richness
of a galaxy cluster as the number of galaxies found within its spatial
boundary and the colour-magnitude filter in the magnitude range $m_{3} < I <
m_{3}+2$, where $m_{3}$ is the magnitude of the third brightest member
galaxy. Although there are some similarities with the Abell richness
definition, the above definition is unique for the FSVS clusters and cannot
be directly translated into Abell richness. According to richness, a subsample of 395 poor ($< 20$ members) clusters has
been separated from the main sample. 117 of the poor clusters are at
relatively high redshift ($z > 0.4$) and due to cosmological dimming only the
brighter part of the $m_{3} < I < m_{3}+2$ range is covered by our data. In
such cases the richness is clearly underestimated.

\begin{figure}
\psfig{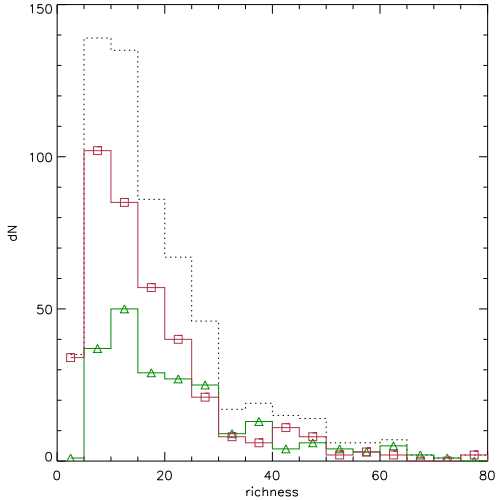}
\caption{The richness distribution of the FSVS cluster sample. For our
purposes, we define the richness of a galaxy cluster as the number of
galaxies found within its spatial boundary and the colour-magnitude filter in
the magnitude range $m_{3} < I < m_{3}+2$, where $m_{3}$ is the magnitude of
the third brightest member galaxy. The dotted histogram corresponds to all
clusters, the solid line with squares to low redshift clusters ($z < 0.4$),
and the solid line with triangles to high redshift clusters ($z \geq 0.4$).}
\label{richness_distr}
\end{figure}

The disadvantage of the calculation of richness from only a relatively small
colour selection is a discrimination against clusters with a high fraction of
blue galaxies that will not contribute to the galaxy count. This then leads
to underestimation of the richness. However, inclusion of all galaxies within
the cluster boundary would require a background substraction. The richness
measure would then become prone to the usually large uncertainty in the
estimate of the counts of background galaxies at the cluster position. A
global background correction, used in the majority of richness estimates, is
not appropriate owing to very strong local variations as found by Valotto,
Moore \& Lambas (2001) and readily visible in the colour-magnitude diagrams
of the example FSVS clusters (Figures~\ref{clust_morph1} and
\ref{clust_morph2}). In both examples the areas of the clusters are very
similar, but, in the first example, the presence of multiple clusters at
different redshifts produces a high local enhancement of the background
galaxies. The richness measure adopted for the FSVS clusters can be
considered as valid for the population of early-type galaxies in the
clusters. Figure~\ref{richness_distr} illustrates the distribution of the
richness of the FSVS clusters, divided into low and high redshift samples
(where ``low'' signifies $z < 0.4$).

\subsection{Redshift distribution}
 
\begin{figure}
\psfig{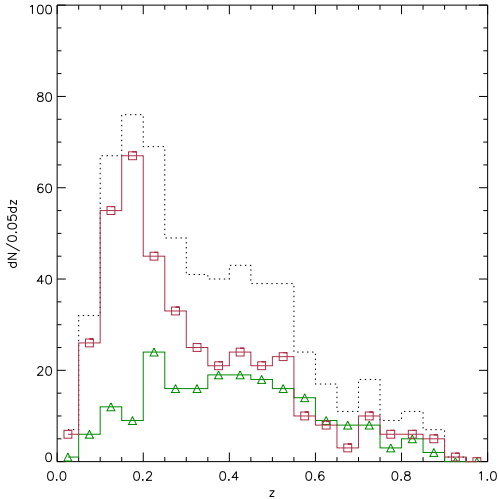}
\caption{The distribution of estimated redshifts in the FSVS cluster
sample. The dotted line histogram corresponds to all clusters, the solid line
with squares to poor clusters (richness $< 20$), and the solid line with
triangles to rich clusters (richness $\geq 20$).}
\label{redshift_distr}
\end{figure}

The 598 clusters detected in $21.75$ deg$^{2}$ of FSVS data correspond to a
mean cluster density of $\sim 28$ deg$^{-2}$ and have a mean redshift of
$\langle z \rangle = 0.345$ (Figure~\ref{redshift_distr}). To compare our
detection rate with, for example, those reported for SDSS data, which has a
different depth, we need to consider the properties of the cluster luminosity
function. The population of bright cluster galaxies (BCGs) is evident in the
cluster luminosity function as a "bump" at bright magnitudes that rises above
the luminosity function of other cluster members (Lin \& Mohr 2004; Hensen et
al.\ 2005). As found by Hensen et al.\ (2005), the luminosity function of
low- and intermediate-richness clusters cannot be well described by a
Schechter function when BCGs are included; this function provides acceptable
fits only for the richer systems where the fractional contribution of the
central galaxies to the LF is small. The presence of the BCG population
allows a high rate of cluster identification even in surveys detecting only a
few (brightest) members. Consequently, up to a certain redshift, two surveys
which are different in depth will still allow similar cluster detection
rates.

We compared the cluster density in the FSVS Cluster Catalogue with the
cluster sample derived from the SDSS data using the Cut-and-Enhance method
(Goto et al.\ 2002). At $z < 0.2$ the density in the FSVS sample is enhanced
relative to that from the SDSS data, since FSVS includes poor groups, which
have few bright galaxies excluding them from the shallower limits of the
SDSS. However, in the restricted redshift range $0.2 < z < 0.3$, the cluster
density of $6.3$ deg$^{-2}$ from FSVS is comparable to the $6.2$ deg$^{-2}$
found from SDSS data. In this transitional redshift range, the FSVS no longer
allows the detection of poor groups but SDSS still has sufficient depth to
see enough bright galaxies for positive cluster identification. At higher
redshifts ($z > 0.3$), the shallower SDSS begins to miss the galaxies in the
bright tail of the luminosity function leading to a rapid decline in the
cluster detection rate. Consequently, to carry out a comparison of both
samples at $z > 0.3$, the magnitude limit of the FSVS data would have to be
restricted to match the SDSS. Instead, we deduce from the results at $z <
0.3$ that both algorithms have similar detection efficiency.

\begin{figure}
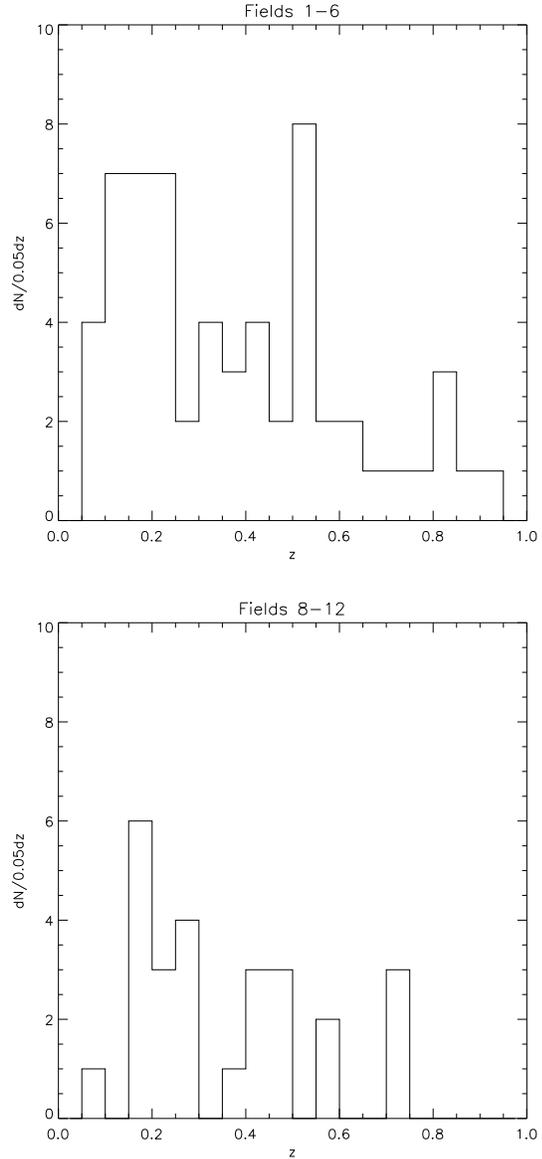

\psfig{file=figure14a.epsi,width=7.0cm}
\vspace{0.5cm}
\psfig{file=figure14b.epsi,width=7.0cm}
\caption{The estimated redshift distribution of galaxy clusters in Fields
1--6 (top) and 8--12 (bottom).}
\label{z_fileds}
\end{figure}

The number of detected clusters $N_{cl}$ varies from field to field, ranging
from 0 to 21. A small part of this variation is contributed by the difference
in the limiting magnitudes arising from changes in observing
conditions. However, the main contribution is due to the presence of voids
and superclusters in the spatial distribution of galaxy clusters, also
deduced from peaks in their redshift distribution if determined field by
field. Figure~\ref{z_fileds} illustrates two examples of groupings of FSVS
fields with prominent peaks in the redshift distribution of clusters,
suggesting the presence of filament- or sheet-like structures.

\subsection{Contamination by spurious clusters}

Chance projections of galaxies similar in colour could cause some spurious
detections in the cluster catalogue. A visual examination of the images
cannot provide a reliable and repeatable check of the physical reality of the
detected clusters. To avoid ambiguity, the reality of a cluster is judged by
the presence of both a local overdensity and a well-defined linear CRS. 

The partial maximised likelihood of a cluster with a positive contrast above
background is

\begin{equation}
l(\bmath{x}; {\mathcal{A}}) > 1.0
\end{equation}

Adopting this value as a threshold to reject spurious clusters, we find a
contamination level of $\sim 15\%$ distributed across the whole redshift
range.

We then use a chi-square statistic as a measure of whether the linear fit to
the colour-magnitude distribution of cluster members is acceptable --- that
is, whether we can consider the presence of a CRS to be believable.

We fit a linear relation to the putative CRS by minimising
\begin{equation}
\chi^{2} = \sum_{i=1}^{n}(y_{i} - (A+Bx_{i}))^{2}/\sigma_{i}^{2}
\end{equation}
where $A$ and $B$ are the coefficients of the linear fit, $n$ is the number
of cluster members and $\sigma_i$ is adopted as the photometric error of the
galaxy colour. The values of A and B are extracted directly from the standard
minimum squares fit without apriori restriction. We shall denote the
minimised value of $\chi^{2}$ that results as $\chi^{2}_{CRS}$.

We then define a ``CRS probability'' as $P_{CRS} = {\rm Pr}(\chi^{2} <
\chi^{2}_{CRS})$ for a $\chi^{2}$ distribution with $n-3$ degrees of freedom.

\begin{figure}
\psfig{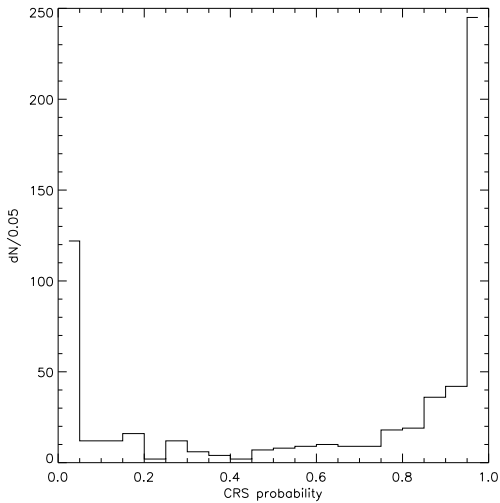}
\caption{Distribution of the ``CRS probability,'' $P_{CRS} = {\rm Pr}
(\chi^{2} < \chi^{2}_{CRS})$ for a $\chi^{2}$ distribution with $n-3$ degrees
of freedom. The value $P_{CRS} \sim 0.4$ at which a miminum is seen is
adopted as the threshold for a believable CRS.}
\label{prob_distr}
\end{figure}

We shall take values of this CRS probability above a specified threshold to
indicate that the presence of a particular CRS is ``believable.'' The
observed distribution of the values of the CRS probabilities is shown in
Figure~\ref{prob_distr}. There is a minimum in the distribution at $P_{CRS}
\sim 0.4$, which, unlike the local minimum at $P_{CRS} \sim 0.2$, appears to
genuinely mark a transition from generally decreasing to generally
increasing. We cautiously but plausibly interpret the transition as being
from predominantly spurious clusters ($P_{CRS} < 0.4$) to predominantly real
clusters ($P_{CRS} \ge 0.4$). We have therefore adopted the value of $P_{CRS}
= 0.4$ as the threshold.

Clearly, this $\chi^{2}$ procedure is not statistically rigorous because the
threshold of $P_{CRS} = 0.4$ is set by judgement and also because of bias to
the $\chi^{2}$-minimisation and calculation of $P_{CRS}$ that results from
the imposition of the colour-magnitude filter. Nevertheless, the procedure
provides a clearly-defined criterion for assessing whether a CRS is
believable. We might find in future that the chosen threshold of $0.4$ will
need some revision.

\begin{figure}
\psfig{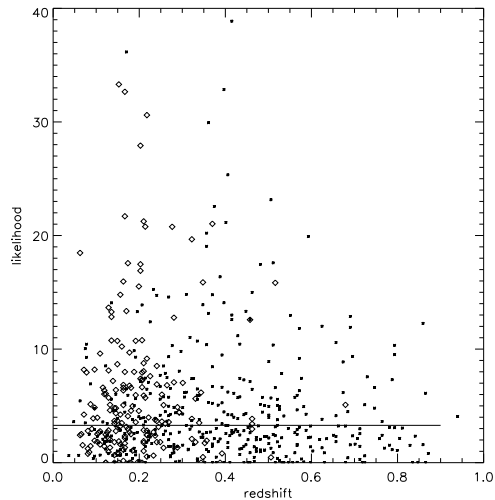}
\caption{Plot of cluster spatial likelihood $l(\bmath{x}; {\mathcal{A}})$
versus redshift. Diamonds indicate clusters with an ``unbelievable'' CRS (CRS
probability $P_{CRS} < 0.4$); crosses mark the remaining cluster
population. The horizontal solid line indicates the median likelihood.}
\label{z_contam}
\end{figure}

Considering the lack of a reliable CRS to be an indication that a cluster is
spurious, the contamination of the sample has been estimated to be $\sim
31\%$. Nevertheless, the level of contamination varies strongly with
redshift. Figure~\ref{z_contam} shows that very few of the higher redshift
($z > 0.4$) clusters are suspected of being spurious. Furthermore, many of
the rejected clusters have a very high spatial likelihood (very high values
of $l(\bmath{x}; {\mathcal{A}})$) suggesting that our test of the presence of
a linear CRS is biased against rich, low-redshift clusters. The
$\chi^{2}_{CRS}$ of these clusters is inflated by a large population of faint
member galaxies occupying much broader colour-space. To overcome this
limitation, only clusters with an ``unbelievable'' CRS combined with
likelihood $l(\bmath{x}; {\mathcal{A}})$ below the median are rejected. This
parameter combination (CRS probability $< 0.4$ and $l(\bmath{x};
{\mathcal{A}}) < 3.28$) rejects $12.8\%$ of clusters as spurious.

\begin{figure}
\psfig{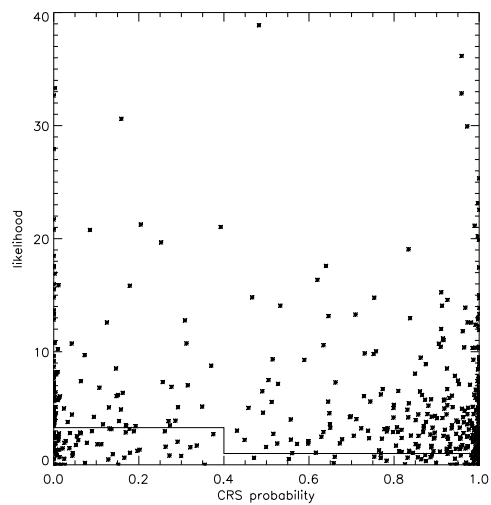}
\caption{Plot of cluster spatial likelihood $l(\bmath{x}; {\mathcal{A}})$
versus CRS probability. All clusters below the solid line are rejected as
spurious.}
\label{spurious_clst}
\end{figure}

Figure~\ref{spurious_clst} illustrates the overall rejection region that is
defined by the restrictions on spatial likelihood and CRS probability
($P_{CRS} < 0.4$ and $l(\bmath{x}; {\mathcal{A}}) < 3.28$, or $l(\bmath{x};
{\mathcal{A}}) < 1.0$). The overall contamination by spurious clusters is
then estimated to be below $24\%$.

The FSVS cluster catalogue is dominated by poor clusters and groups that may
lack central bright galaxies and that show small diversity in magnitude.
Consequently, many of the clusters found to lack a strong CRS may still be
real and the estimated contamination has to be considered to be an upper
limit.

\section{The Catalogue}

A wide range of parameters has been derived for every detected cluster. The
catalogue given here, Table~\ref{catalogue}, contains only the most basic
information such as ID, position and estimated redshift. 

\begin{table*}
\center
\caption{The basic FSVS cluster catalogue. The columns are: (1)
identification of the cluster; (2) and (3) RA, Dec (J2000) of the cluster
centre, defined as the mean position of its members; (4) estimated redshift,
$z_{est}$.}
\label{catalogue}
\begin{tabular}{llllllll}
\hline
ID  & RA(J2000)  &  Dec(J2000)  &  $z_{est}$  & ID  & RA(J2000)  &  Dec(J2000)  &  $z_{est}$ \\
\hline
FSVS$\_$CL234208$+$265328 & 23:42:08 & 26:53:28 &  0.13 & FSVS$\_$CL234251$+$270509 & 23:42:51 & 27:05:09 &  0.15 \\
FSVS$\_$CL234146$+$265554 & 23:41:46 & 26:55:54 &  0.21 & FSVS$\_$CL234208$+$270501 & 23:42:08 & 27:05:01 &  0.22 \\
FSVS$\_$CL234201$+$271109 & 23:42:01 & 27:11:09 &  0.39 & FSVS$\_$CL234057$+$270648 & 23:40:57 & 27:06:48 &  0.53 \\
FSVS$\_$CL234208$+$270739 & 23:42:08 & 27:07:39 &  0.66 & FSVS$\_$CL234150$+$265516 & 23:41:50 & 26:55:16 &  0.94 \\
FSVS$\_$CL234326$+$270343 & 23:43:26 & 27:03:43 &  0.11 & FSVS$\_$CL234404$+$270134 & 23:44:04 & 27:01:34 &  0.11 \\
FSVS$\_$CL234439$+$270201 & 23:44:39 & 27:02:01 &  0.21 & FSVS$\_$CL234410$+$270022 & 23:44:10 & 27:00:22 &  0.26 \\
FSVS$\_$CL234505$+$265235 & 23:45:05 & 26:52:35 &  0.21 & FSVS$\_$CL234441$+$270217 & 23:44:41 & 27:02:17 &  0.37 \\
FSVS$\_$CL234448$+$270754 & 23:44:48 & 27:07:54 &  0.45 & FSVS$\_$CL234500$+$265648 & 23:45:00 & 26:56:48 &  0.52 \\
FSVS$\_$CL234455$+$265542 & 23:44:55 & 26:55:42 &  0.58 & FSVS$\_$CL234449$+$270551 & 23:44:49 & 27:05:51 &  0.52 \\
FSVS$\_$CL234439$+$270420 & 23:44:39 & 27:04:20 &  0.79 & FSVS$\_$CL234624$+$265637 & 23:46:24 & 26:56:37 &  0.08 \\
FSVS$\_$CL234610$+$270123 & 23:46:10 & 27:01:23 &  0.15 & FSVS$\_$CL234621$+$265307 & 23:46:21 & 26:53:07 &  0.07 \\
FSVS$\_$CL234600$+$271757 & 23:46:00 & 27:17:57 &  0.16 & FSVS$\_$CL234646$+$270852 & 23:46:46 & 27:08:52 &  0.14 \\
FSVS$\_$CL234614$+$265034 & 23:46:14 & 26:50:34 &  0.19 & FSVS$\_$CL234540$+$265115 & 23:45:40 & 26:51:15 &  0.27 \\
FSVS$\_$CL234607$+$265222 & 23:46:07 & 26:52:22 &  0.19 & FSVS$\_$CL234658$+$270857 & 23:46:58 & 27:08:57 &  0.22 \\
FSVS$\_$CL234607$+$265323 & 23:46:07 & 26:53:23 &  0.51 & FSVS$\_$CL234621$+$270202 & 23:46:21 & 27:02:02 &  0.33 \\
FSVS$\_$CL234613$+$265333 & 23:46:13 & 26:53:33 &  0.35 & FSVS$\_$CL234631$+$265834 & 23:46:31 & 26:58:34 &  0.38 \\
FSVS$\_$CL234445$+$270756 & 23:44:45 & 27:07:56 &  0.40 & FSVS$\_$CL234545$+$270456 & 23:45:45 & 27:04:56 &  0.54 \\
FSVS$\_$CL234537$+$270728 & 23:45:37 & 27:07:28 &  0.70 & FSVS$\_$CL234648$+$273728 & 23:46:48 & 27:37:28 &  0.24 \\
FSVS$\_$CL234610$+$272345 & 23:46:10 & 27:23:45 &  0.29 & FSVS$\_$CL234635$+$275013 & 23:46:35 & 27:50:13 &  0.33 \\
FSVS$\_$CL234641$+$273702 & 23:46:41 & 27:37:02 &  0.33 & FSVS$\_$CL234607$+$272123 & 23:46:07 & 27:21:23 &  0.49 \\
FSVS$\_$CL234604$+$272231 & 23:46:04 & 27:22:31 &  0.55 & FSVS$\_$CL234559$+$273049 & 23:45:59 & 27:30:49 &  0.62 \\
FSVS$\_$CL234557$+$272616 & 23:45:57 & 27:26:16 &  0.84 & FSVS$\_$CL234639$+$275029 & 23:46:39 & 27:50:29 &  0.80 \\
FSVS$\_$CL234430$+$274709 & 23:44:30 & 27:47:09 &  0.14 & FSVS$\_$CL234417$+$272714 & 23:44:17 & 27:27:14 &  0.14 \\
FSVS$\_$CL234456$+$272255 & 23:44:56 & 27:22:55 &  0.17 & FSVS$\_$CL234437$+$274550 & 23:44:37 & 27:45:50 &  0.19 \\
FSVS$\_$CL234337$+$273622 & 23:43:37 & 27:36:22 &  0.40 & FSVS$\_$CL234453$+$272254 & 23:44:53 & 27:22:54 &  0.43 \\
FSVS$\_$CL234312$+$272327 & 23:43:12 & 27:23:27 &  0.51 & FSVS$\_$CL234351$+$272825 & 23:43:51 & 27:28:25 &  0.79 \\
FSVS$\_$CL234138$+$273954 & 23:41:38 & 27:39:54 &  0.08 & FSVS$\_$CL234055$+$273202 & 23:40:55 & 27:32:02 &  0.10 \\
FSVS$\_$CL234100$+$272828 & 23:41:00 & 27:28:28 &  0.22 & FSVS$\_$CL234134$+$271950 & 23:41:34 & 27:19:50 &  0.17 \\
FSVS$\_$CL234247$+$272400 & 23:42:47 & 27:24:00 &  0.46 & FSVS$\_$CL234144$+$273214 & 23:41:44 & 27:32:14 &  0.49 \\
FSVS$\_$CL234141$+$273135 & 23:41:41 & 27:31:35 &  0.53 & FSVS$\_$CL234137$+$273143 & 23:41:37 & 27:31:43 &  0.79 \\
FSVS$\_$CL022856$+$144800 & 02:28:56 & 14:48:00 &  0.20 & FSVS$\_$CL022956$+$145701 & 02:29:56 & 14:57:01 &  0.27 \\
FSVS$\_$CL022931$+$143738 & 02:29:31 & 14:37:38 &  0.27 & FSVS$\_$CL022925$+$144632 & 02:29:25 & 14:46:32 &  0.27 \\
FSVS$\_$CL022855$+$144801 & 02:28:55 & 14:48:01 &  0.42 & FSVS$\_$CL022813$+$144547 & 02:28:13 & 14:45:47 &  0.57 \\
FSVS$\_$CL022844$+$144712 & 02:28:44 & 14:47:12 &  0.56 & FSVS$\_$CL022947$+$143524 & 02:29:47 & 14:35:24 &  0.72 \\
FSVS$\_$CL023117$+$150515 & 02:31:17 & 15:05:15 &  0.18 & FSVS$\_$CL023101$+$144945 & 02:31:01 & 14:49:45 &  0.19 \\
FSVS$\_$CL023057$+$144500 & 02:30:57 & 14:45:00 &  0.38 & FSVS$\_$CL023001$+$145656 & 02:30:01 & 14:56:56 &  0.41 \\
FSVS$\_$CL022957$+$145747 & 02:29:57 & 14:57:47 &  0.48 & FSVS$\_$CL023509$+$150306 & 02:35:09 & 15:03:06 &  0.07 \\
FSVS$\_$CL023420$+$151633 & 02:34:20 & 15:16:33 &  0.16 & FSVS$\_$CL023412$+$150004 & 02:34:12 & 15:00:04 &  0.18 \\
FSVS$\_$CL023341$+$151137 & 02:33:41 & 15:11:37 &  0.39 & FSVS$\_$CL023332$+$145904 & 02:33:32 & 14:59:04 &  0.47 \\
FSVS$\_$CL023432$+$151326 & 02:34:32 & 15:13:26 &  0.72 & FSVS$\_$CL023647$+$152254 & 02:36:47 & 15:22:54 &  0.18 \\
FSVS$\_$CL023643$+$152354 & 02:36:43 & 15:23:54 &  0.26 & FSVS$\_$CL023621$+$151025 & 02:36:21 & 15:10:25 &  0.44 \\
FSVS$\_$CL023635$+$151533 & 02:36:35 & 15:15:33 &  0.69 & FSVS$\_$CL023845$+$153632 & 02:38:45 & 15:36:32 &  0.19 \\
FSVS$\_$CL023832$+$153907 & 02:38:32 & 15:39:07 &  0.21 & FSVS$\_$CL023816$+$152247 & 02:38:16 & 15:22:47 &  0.22 \\
FSVS$\_$CL074721$+$205219 & 07:47:21 & 20:52:19 &  0.14 & FSVS$\_$CL074808$+$210159 & 07:48:08 & 21:01:59 &  0.14 \\
FSVS$\_$CL074813$+$205529 & 07:48:13 & 20:55:29 &  0.18 & FSVS$\_$CL074807$+$210550 & 07:48:07 & 21:05:50 &  0.27 \\
FSVS$\_$CL074725$+$204939 & 07:47:25 & 20:49:39 &  0.28 & FSVS$\_$CL075014$+$205839 & 07:50:14 & 20:58:39 &  0.08 \\
FSVS$\_$CL074922$+$204039 & 07:49:22 & 20:40:39 &  0.16 & FSVS$\_$CL075053$+$205247 & 07:50:53 & 20:52:47 &  0.16 \\
FSVS$\_$CL075100$+$204842 & 07:51:00 & 20:48:42 &  0.19 & FSVS$\_$CL075034$+$205510 & 07:50:34 & 20:55:10 &  0.24 \\
FSVS$\_$CL075032$+$205545 & 07:50:32 & 20:55:45 &  0.36 & FSVS$\_$CL075055$+$210542 & 07:50:55 & 21:05:42 &  0.36 \\
FSVS$\_$CL075053$+$205221 & 07:50:53 & 20:52:21 &  0.51 & FSVS$\_$CL075155$+$203343 & 07:51:55 & 20:33:43 &  0.08 \\
FSVS$\_$CL075148$+$204634 & 07:51:48 & 20:46:34 &  0.09 & FSVS$\_$CL075307$+$204636 & 07:53:07 & 20:46:36 &  0.18 \\
FSVS$\_$CL075220$+$204328 & 07:52:20 & 20:43:28 &  0.14 & FSVS$\_$CL075254$+$203123 & 07:52:54 & 20:31:23 &  0.27 \\
FSVS$\_$CL075314$+$203637 & 07:53:14 & 20:36:37 &  0.31 & FSVS$\_$CL075202$+$204352 & 07:52:02 & 20:43:52 &  0.44 \\
FSVS$\_$CL075455$+$203646 & 07:54:55 & 20:36:46 &  0.09 & FSVS$\_$CL075348$+$204158 & 07:53:48 & 20:41:58 &  0.15 \\
FSVS$\_$CL075406$+$203207 & 07:54:06 & 20:32:07 &  0.18 & FSVS$\_$CL075415$+$203426 & 07:54:15 & 20:34:26 &  0.20 \\
FSVS$\_$CL075529$+$204532 & 07:55:29 & 20:45:32 &  0.29 & FSVS$\_$CL075504$+$204520 & 07:55:04 & 20:45:20 &  0.44 \\
FSVS$\_$CL075430$+$202712 & 07:54:30 & 20:27:12 &  0.57 & FSVS$\_$CL075712$+$202015 & 07:57:12 & 20:20:15 &  0.19 \\
\hline
\end{tabular}
\end{table*}

\addtocounter{table}{-1}
\begin{table*}
\center
\caption{Continued.}
\begin{tabular}{llllllll}
\hline
ID  & RA(J2000)  &  Dec(J2000)  &  $z_{est}$  & ID  & RA(J2000)  &  Dec(J2000)  &  $z_{est}$ \\
\hline
FSVS$\_$CL075656$+$203049 & 07:56:56 & 20:30:49 &  0.41 & FSVS$\_$CL075702$+$203421 & 07:57:02 & 20:34:21 &  0.86 \\
FSVS$\_$CL075908$+$205955 & 07:59:08 & 20:59:55 &  0.07 & FSVS$\_$CL075842$+$210256 & 07:58:42 & 21:02:56 &  0.14 \\
FSVS$\_$CL075853$+$211245 & 07:58:53 & 21:12:45 &  0.20 & FSVS$\_$CL075846$+$205104 & 07:58:46 & 20:51:04 &  0.22 \\
FSVS$\_$CL075801$+$205055 & 07:58:01 & 20:50:55 &  0.25 & FSVS$\_$CL075858$+$210527 & 07:58:58 & 21:05:27 &  0.22 \\
FSVS$\_$CL075828$+$210023 & 07:58:28 & 21:00:23 &  0.31 & FSVS$\_$CL075807$+$205146 & 07:58:07 & 20:51:46 &  0.43 \\
FSVS$\_$CL075808$+$211321 & 07:58:08 & 21:13:21 &  0.52 & FSVS$\_$CL125049$+$264834 & 12:50:49 & 26:48:34 &  0.41 \\
FSVS$\_$CL125006$+$270117 & 12:50:06 & 27:01:17 &  0.12 & FSVS$\_$CL124938$+$265301 & 12:49:38 & 26:53:01 &  0.14 \\
FSVS$\_$CL124915$+$264620 & 12:49:15 & 26:46:20 &  0.30 & FSVS$\_$CL124917$+$271153 & 12:49:17 & 27:11:53 &  0.28 \\
FSVS$\_$CL124955$+$264645 & 12:49:55 & 26:46:45 &  0.37 & FSVS$\_$CL124842$+$270735 & 12:48:42 & 27:07:35 &  0.44 \\
FSVS$\_$CL124845$+$265335 & 12:48:45 & 26:53:35 &  0.52 & FSVS$\_$CL124854$+$271449 & 12:48:54 & 27:14:49 &  0.52 \\
FSVS$\_$CL124903$+$271259 & 12:49:03 & 27:12:59 &  0.57 & FSVS$\_$CL124918$+$271135 & 12:49:18 & 27:11:35 &  0.71 \\
FSVS$\_$CL125315$+$271527 & 12:53:15 & 27:15:27 &  0.25 & FSVS$\_$CL125357$+$265044 & 12:53:57 & 26:50:44 &  0.31 \\
FSVS$\_$CL125258$+$270556 & 12:52:58 & 27:05:56 &  0.45 & FSVS$\_$CL125332$+$271332 & 12:53:32 & 27:13:32 &  0.42 \\
FSVS$\_$CL125312$+$265349 & 12:53:12 & 26:53:49 &  0.45 & FSVS$\_$CL125300$+$272116 & 12:53:00 & 27:21:16 &  0.45 \\
FSVS$\_$CL125539$+$265122 & 12:55:39 & 26:51:22 &  0.12 & FSVS$\_$CL125511$+$265510 & 12:55:11 & 26:55:10 &  0.10 \\
FSVS$\_$CL125627$+$264800 & 12:56:27 & 26:48:00 &  0.09 & FSVS$\_$CL125524$+$265800 & 12:55:24 & 26:58:00 &  0.41 \\
FSVS$\_$CL125535$+$264549 & 12:55:35 & 26:45:49 &  0.17 & FSVS$\_$CL125602$+$265121 & 12:56:02 & 26:51:21 &  0.12 \\
FSVS$\_$CL125540$+$264517 & 12:55:40 & 26:45:17 &  0.16 & FSVS$\_$CL125535$+$270036 & 12:55:35 & 27:00:36 &  0.27 \\
FSVS$\_$CL125423$+$265107 & 12:54:23 & 26:51:07 &  0.24 & FSVS$\_$CL125636$+$270837 & 12:56:36 & 27:08:37 &  0.25 \\
FSVS$\_$CL125611$+$265756 & 12:56:11 & 26:57:56 &  0.27 & FSVS$\_$CL125550$+$270450 & 12:55:50 & 27:04:50 &  0.36 \\
FSVS$\_$CL125555$+$271626 & 12:55:55 & 27:16:26 &  0.41 & FSVS$\_$CL125531$+$265746 & 12:55:31 & 26:57:46 &  0.59 \\
FSVS$\_$CL125628$+$265333 & 12:56:28 & 26:53:33 &  0.86 & FSVS$\_$CL125428$+$261050 & 12:54:28 & 26:10:50 &  0.15 \\
FSVS$\_$CL125520$+$261912 & 12:55:20 & 26:19:12 &  0.15 & FSVS$\_$CL125515$+$261842 & 12:55:15 & 26:18:42 &  0.22 \\
FSVS$\_$CL125559$+$260443 & 12:55:59 & 26:04:43 &  0.18 & FSVS$\_$CL125544$+$262518 & 12:55:44 & 26:25:18 &  0.15 \\
FSVS$\_$CL125632$+$262418 & 12:56:32 & 26:24:18 &  0.19 & FSVS$\_$CL125634$+$260833 & 12:56:34 & 26:08:33 &  0.17 \\
FSVS$\_$CL125438$+$261716 & 12:54:38 & 26:17:16 &  0.16 & FSVS$\_$CL125513$+$262023 & 12:55:13 & 26:20:23 &  0.31 \\
FSVS$\_$CL125537$+$262409 & 12:55:37 & 26:24:09 &  0.29 & FSVS$\_$CL125539$+$262545 & 12:55:39 & 26:25:45 &  0.39 \\
FSVS$\_$CL125442$+$262109 & 12:54:42 & 26:21:09 &  0.46 & FSVS$\_$CL125556$+$260810 & 12:55:56 & 26:08:10 &  0.60 \\
FSVS$\_$CL125617$+$262337 & 12:56:17 & 26:23:37 &  0.53 & FSVS$\_$CL125335$+$261534 & 12:53:35 & 26:15:34 &  0.07 \\
FSVS$\_$CL125332$+$261305 & 12:53:32 & 26:13:05 &  0.09 & FSVS$\_$CL125418$+$260931 & 12:54:18 & 26:09:31 &  0.06 \\
FSVS$\_$CL125322$+$262508 & 12:53:22 & 26:25:08 &  0.13 & FSVS$\_$CL125358$+$261956 & 12:53:58 & 26:19:56 &  0.12 \\
FSVS$\_$CL125327$+$263249 & 12:53:27 & 26:32:49 &  0.12 & FSVS$\_$CL125317$+$263249 & 12:53:17 & 26:32:49 &  0.15 \\
FSVS$\_$CL125411$+$261259 & 12:54:11 & 26:12:59 &  0.18 & FSVS$\_$CL125222$+$261612 & 12:52:22 & 26:16:12 &  0.26 \\
FSVS$\_$CL125316$+$260912 & 12:53:16 & 26:09:12 &  0.24 & FSVS$\_$CL125255$+$261808 & 12:52:55 & 26:18:08 &  0.36 \\
FSVS$\_$CL125353$+$261739 & 12:53:53 & 26:17:39 &  0.30 & FSVS$\_$CL125312$+$261217 & 12:53:12 & 26:12:17 &  0.52 \\
FSVS$\_$CL125318$+$260652 & 12:53:18 & 26:06:52 &  0.52 & FSVS$\_$CL125326$+$262052 & 12:53:26 & 26:20:52 &  0.51 \\
FSVS$\_$CL125307$+$261714 & 12:53:07 & 26:17:14 &  0.67 & FSVS$\_$CL125039$+$261548 & 12:50:39 & 26:15:48 &  0.18 \\
FSVS$\_$CL125143$+$263329 & 12:51:43 & 26:33:29 &  0.12 & FSVS$\_$CL125150$+$263419 & 12:51:50 & 26:34:19 &  0.11 \\
FSVS$\_$CL124958$+$262347 & 12:49:58 & 26:23:47 &  0.14 & FSVS$\_$CL125028$+$261838 & 12:50:28 & 26:18:38 &  0.21 \\
FSVS$\_$CL125017$+$262617 & 12:50:17 & 26:26:17 &  0.12 & FSVS$\_$CL125050$+$262435 & 12:50:50 & 26:24:35 &  0.20 \\
FSVS$\_$CL124955$+$261746 & 12:49:55 & 26:17:46 &  0.20 & FSVS$\_$CL125012$+$261104 & 12:50:12 & 26:11:04 &  0.25 \\
FSVS$\_$CL125109$+$261151 & 12:51:09 & 26:11:51 &  0.27 & FSVS$\_$CL125048$+$263418 & 12:50:48 & 26:34:18 &  0.30 \\
FSVS$\_$CL125035$+$261332 & 12:50:35 & 26:13:32 &  0.38 & FSVS$\_$CL125006$+$260853 & 12:50:06 & 26:08:53 &  0.37 \\
FSVS$\_$CL125030$+$262743 & 12:50:30 & 26:27:43 &  0.33 & FSVS$\_$CL125009$+$260931 & 12:50:09 & 26:09:31 &  0.41 \\
FSVS$\_$CL125145$+$263824 & 12:51:45 & 26:38:24 &  0.37 & FSVS$\_$CL125030$+$262535 & 12:50:30 & 26:25:35 &  0.46 \\
FSVS$\_$CL125007$+$261031 & 12:50:07 & 26:10:31 &  0.45 & FSVS$\_$CL125140$+$262640 & 12:51:40 & 26:26:40 &  0.62 \\
FSVS$\_$CL125028$+$262052 & 12:50:28 & 26:20:52 &  0.73 & FSVS$\_$CL124929$+$255843 & 12:49:29 & 25:58:43 &  0.16 \\
FSVS$\_$CL124755$+$255527 & 12:47:55 & 25:55:27 &  0.08 & FSVS$\_$CL124925$+$260755 & 12:49:25 & 26:07:55 &  0.14 \\
FSVS$\_$CL124904$+$255734 & 12:49:04 & 25:57:34 &  0.28 & FSVS$\_$CL124943$+$261754 & 12:49:43 & 26:17:54 &  0.53 \\
FSVS$\_$CL162235$+$263052 & 16:22:35 & 26:30:52 &  0.17 & FSVS$\_$CL162421$+$264423 & 16:24:21 & 26:44:23 &  0.12 \\
FSVS$\_$CL162424$+$262315 & 16:24:24 & 26:23:15 &  0.14 & FSVS$\_$CL162339$+$263433 & 16:23:39 & 26:34:33 &  0.39 \\
FSVS$\_$CL162415$+$263854 & 16:24:15 & 26:38:54 &  0.15 & FSVS$\_$CL162412$+$263008 & 16:24:12 & 26:30:08 &  0.37 \\
FSVS$\_$CL162358$+$264914 & 16:23:58 & 26:49:14 &  0.71 & FSVS$\_$CL162420$+$263719 & 16:24:20 & 26:37:19 &  0.63 \\
FSVS$\_$CL162434$+$263800 & 16:24:34 & 26:38:00 &  0.80 & FSVS$\_$CL162154$+$262742 & 16:21:54 & 26:27:42 &  0.10 \\
FSVS$\_$CL162229$+$264730 & 16:22:29 & 26:47:30 &  0.16 & FSVS$\_$CL162131$+$262758 & 16:21:31 & 26:27:58 &  0.13 \\
FSVS$\_$CL162231$+$262657 & 16:22:31 & 26:26:57 &  0.18 & FSVS$\_$CL162045$+$262549 & 16:20:45 & 26:25:49 &  0.12 \\
FSVS$\_$CL162200$+$262722 & 16:22:00 & 26:27:22 &  0.25 & FSVS$\_$CL162108$+$262732 & 16:21:08 & 26:27:32 &  0.28 \\
FSVS$\_$CL162119$+$264758 & 16:21:19 & 26:47:58 &  0.68 & FSVS$\_$CL162508$+$263444 & 16:25:08 & 26:34:44 &  0.06 \\
FSVS$\_$CL162518$+$262857 & 16:25:18 & 26:28:57 &  0.11 & FSVS$\_$CL162510$+$262634 & 16:25:10 & 26:26:34 &  0.24 \\
FSVS$\_$CL162515$+$262418 & 16:25:15 & 26:24:18 &  0.37 & FSVS$\_$CL162520$+$262409 & 16:25:20 & 26:24:09 &  0.40 \\
\hline								      
\end{tabular}
\end{table*}

\addtocounter{table}{-1}
\begin{table*}
\center
\caption{Continued.}
\begin{tabular}{llllllll}
\hline
ID  & RA(J2000)  &  Dec(J2000)  &  $z_{est}$  & ID  & RA(J2000)  &  Dec(J2000)  &  $z_{est}$ \\
\hline
FSVS$\_$CL162855$+$263430 & 16:28:55 & 26:34:30 &  0.13 & FSVS$\_$CL162832$+$263640 & 16:28:32 & 26:36:40 &  0.28 \\
FSVS$\_$CL162849$+$263741 & 16:28:49 & 26:37:41 &  0.31 & FSVS$\_$CL162810$+$263537 & 16:28:10 & 26:35:37 &  0.35 \\
FSVS$\_$CL162706$+$263339 & 16:27:06 & 26:33:39 &  0.36 & FSVS$\_$CL172015$+$263858 & 17:20:15 & 26:38:58 &  0.10 \\
FSVS$\_$CL172110$+$262919 & 17:21:10 & 26:29:19 &  0.17 & FSVS$\_$CL172059$+$265741 & 17:20:59 & 26:57:41 &  0.13 \\
FSVS$\_$CL171930$+$263253 & 17:19:30 & 26:32:53 &  0.22 & FSVS$\_$CL172107$+$265336 & 17:21:07 & 26:53:36 &  0.27 \\
FSVS$\_$CL172008$+$264112 & 17:20:08 & 26:41:12 &  0.35 & FSVS$\_$CL171930$+$263812 & 17:19:30 & 26:38:12 &  0.39 \\
FSVS$\_$CL171914$+$264335 & 17:19:14 & 26:43:35 &  0.49 & FSVS$\_$CL171932$+$262948 & 17:19:32 & 26:29:48 &  0.50 \\
FSVS$\_$CL172030$+$261200 & 17:20:30 & 26:12:00 &  0.09 & FSVS$\_$CL172004$+$261810 & 17:20:04 & 26:18:10 &  0.03 \\
FSVS$\_$CL171901$+$260202 & 17:19:01 & 26:02:02 &  0.11 & FSVS$\_$CL172017$+$255048 & 17:20:17 & 25:50:48 &  0.20 \\
FSVS$\_$CL172027$+$261323 & 17:20:27 & 26:13:23 &  0.13 & FSVS$\_$CL171852$+$255236 & 17:18:52 & 25:52:36 &  0.18 \\
FSVS$\_$CL172020$+$260951 & 17:20:20 & 26:09:51 &  0.17 & FSVS$\_$CL171853$+$255038 & 17:18:53 & 25:50:38 &  0.30 \\
FSVS$\_$CL172046$+$261308 & 17:20:46 & 26:13:08 &  0.22 & FSVS$\_$CL171930$+$260325 & 17:19:30 & 26:03:25 &  0.28 \\
FSVS$\_$CL172042$+$260436 & 17:20:42 & 26:04:36 &  0.32 & FSVS$\_$CL171931$+$261901 & 17:19:31 & 26:19:01 &  0.36 \\
FSVS$\_$CL171849$+$255300 & 17:18:49 & 25:53:00 &  0.48 & FSVS$\_$CL172048$+$254819 & 17:20:48 & 25:48:19 &  0.55 \\
FSVS$\_$CL171857$+$255213 & 17:18:57 & 25:52:13 &  0.84 & FSVS$\_$CL172053$+$272314 & 17:20:53 & 27:23:14 &  0.06 \\
FSVS$\_$CL172037$+$271636 & 17:20:37 & 27:16:36 &  0.09 & FSVS$\_$CL172032$+$271607 & 17:20:32 & 27:16:07 &  0.20 \\
FSVS$\_$CL172109$+$265947 & 17:21:09 & 26:59:47 &  0.18 & FSVS$\_$CL172105$+$271911 & 17:21:05 & 27:19:11 &  0.15 \\
FSVS$\_$CL171934$+$271032 & 17:19:34 & 27:10:32 &  0.22 & FSVS$\_$CL171944$+$271429 & 17:19:44 & 27:14:29 &  0.20 \\
FSVS$\_$CL171945$+$271051 & 17:19:45 & 27:10:51 &  0.39 & FSVS$\_$CL171942$+$270906 & 17:19:42 & 27:09:06 &  0.62 \\
FSVS$\_$CL172010$+$271202 & 17:20:10 & 27:12:02 &  0.78 & FSVS$\_$CL171937$+$270740 & 17:19:37 & 27:07:40 &  0.69 \\
FSVS$\_$CL171944$+$265634 & 17:19:44 & 26:56:34 &  0.87 & FSVS$\_$CL172030$+$274013 & 17:20:30 & 27:40:13 &  0.18 \\
FSVS$\_$CL172100$+$273642 & 17:21:00 & 27:36:42 &  0.12 & FSVS$\_$CL172109$+$274329 & 17:21:09 & 27:43:29 &  0.19 \\
FSVS$\_$CL172123$+$274750 & 17:21:23 & 27:47:50 &  0.21 & FSVS$\_$CL172125$+$274456 & 17:21:25 & 27:44:56 &  0.27 \\
FSVS$\_$CL172124$+$274457 & 17:21:24 & 27:44:57 &  0.33 & FSVS$\_$CL172015$+$273823 & 17:20:15 & 27:38:23 &  0.37 \\
FSVS$\_$CL171936$+$273208 & 17:19:36 & 27:32:08 &  0.52 & FSVS$\_$CL172004$+$273733 & 17:20:04 & 27:37:33 &  0.62 \\
FSVS$\_$CL172003$+$273338 & 17:20:03 & 27:33:38 &  0.85 & FSVS$\_$CL172026$+$274516 & 17:20:26 & 27:45:16 &  0.87 \\
FSVS$\_$CL025812$+$192245 & 02:58:12 & 19:22:45 &  0.39 & FSVS$\_$CL025820$+$193519 & 02:58:20 & 19:35:19 &  0.46 \\
FSVS$\_$CL025730$+$193637 & 02:57:30 & 19:36:37 &  0.47 & FSVS$\_$CL025812$+$193852 & 02:58:12 & 19:38:52 &  0.66 \\
FSVS$\_$CL025817$+$193356 & 02:58:17 & 19:33:56 &  0.68 & FSVS$\_$CL030348$+$193451 & 03:03:48 & 19:34:51 &  0.04 \\
FSVS$\_$CL030327$+$194215 & 03:03:27 & 19:42:15 &  0.09 & FSVS$\_$CL030304$+$193601 & 03:03:04 & 19:36:01 &  0.19 \\
FSVS$\_$CL030327$+$194214 & 03:03:27 & 19:42:14 &  0.22 & FSVS$\_$CL030233$+$193522 & 03:02:33 & 19:35:22 &  0.29 \\
FSVS$\_$CL030258$+$195007 & 03:02:58 & 19:50:07 &  0.47 & FSVS$\_$CL030239$+$192933 & 03:02:39 & 19:29:33 &  0.46 \\
FSVS$\_$CL030345$+$193647 & 03:03:45 & 19:36:47 &  0.44 & FSVS$\_$CL030251$+$192415 & 03:02:51 & 19:24:15 &  0.53 \\
FSVS$\_$CL030244$+$194036 & 03:02:44 & 19:40:36 &  0.46 & FSVS$\_$CL030257$+$192619 & 03:02:57 & 19:26:19 &  0.54 \\
FSVS$\_$CL030537$+$195347 & 03:05:37 & 19:53:47 &  0.11 & FSVS$\_$CL030449$+$193756 & 03:04:49 & 19:37:56 &  0.33 \\
FSVS$\_$CL030508$+$193023 & 03:05:08 & 19:30:23 &  0.41 & FSVS$\_$CL030855$+$194515 & 03:08:55 & 19:45:15 &  0.17 \\
FSVS$\_$CL031105$+$193428 & 03:11:05 & 19:34:28 &  0.12 & FSVS$\_$CL031103$+$194936 & 03:11:03 & 19:49:36 &  0.41 \\
FSVS$\_$CL071453$+$200648 & 07:14:53 & 20:06:48 &  0.07 & FSVS$\_$CL071449$+$200245 & 07:14:49 & 20:02:45 &  0.16 \\
FSVS$\_$CL071503$+$204801 & 07:15:03 & 20:48:01 &  0.53 & FSVS$\_$CL071443$+$210911 & 07:14:43 & 21:09:11 &  0.15 \\
FSVS$\_$CL071458$+$211228 & 07:14:58 & 21:12:28 &  0.15 & FSVS$\_$CL071520$+$211013 & 07:15:20 & 21:10:13 &  0.19 \\
FSVS$\_$CL071402$+$210027 & 07:14:02 & 21:00:27 &  0.47 & FSVS$\_$CL071405$+$205947 & 07:14:05 & 20:59:47 &  0.57 \\
FSVS$\_$CL071519$+$211256 & 07:15:19 & 21:12:56 &  0.59 & FSVS$\_$CL071535$+$212351 & 07:15:35 & 21:23:51 &  0.78 \\
FSVS$\_$CL071444$+$214622 & 07:14:44 & 21:46:22 &  0.12 & FSVS$\_$CL071558$+$214733 & 07:15:58 & 21:47:33 &  0.13 \\
FSVS$\_$CL071550$+$214842 & 07:15:50 & 21:48:42 &  0.16 & FSVS$\_$CL071535$+$214256 & 07:15:35 & 21:42:56 &  0.21 \\
FSVS$\_$CL071520$+$220013 & 07:15:20 & 22:00:13 &  0.28 & FSVS$\_$CL071530$+$214206 & 07:15:30 & 21:42:06 &  0.32 \\
FSVS$\_$CL071436$+$214204 & 07:14:36 & 21:42:04 &  0.29 & FSVS$\_$CL071521$+$214116 & 07:15:21 & 21:41:16 &  0.66 \\
FSVS$\_$CL071405$+$221159 & 07:14:05 & 22:11:59 &  0.20 & FSVS$\_$CL071508$+$223233 & 07:15:08 & 22:32:33 &  0.22 \\
FSVS$\_$CL071402$+$221709 & 07:14:02 & 22:17:09 &  0.19 & FSVS$\_$CL071502$+$223716 & 07:15:02 & 22:37:16 &  0.40 \\
FSVS$\_$CL071431$+$221018 & 07:14:31 & 22:10:18 &  0.48 & FSVS$\_$CL071550$+$223003 & 07:15:50 & 22:30:03 &  0.70 \\
FSVS$\_$CL071516$+$234824 & 07:15:16 & 23:48:24 &  0.12 & FSVS$\_$CL071523$+$234841 & 07:15:23 & 23:48:41 &  0.20 \\
FSVS$\_$CL071422$+$233547 & 07:14:22 & 23:35:47 &  0.28 & FSVS$\_$CL071455$+$232606 & 07:14:55 & 23:26:06 &  0.38 \\
FSVS$\_$CL071357$+$233017 & 07:13:57 & 23:30:17 &  0.66 & FSVS$\_$CL071536$+$232254 & 07:15:36 & 23:22:54 &  0.57 \\
FSVS$\_$CL100031$+$200527 & 10:00:31 & 20:05:27 &  0.13 & FSVS$\_$CL095926$+$205923 & 09:59:26 & 20:59:23 &  0.16 \\
FSVS$\_$CL100005$+$211017 & 10:00:05 & 21:10:17 &  0.21 & FSVS$\_$CL100034$+$212432 & 10:00:34 & 21:24:32 &  0.24 \\
FSVS$\_$CL100007$+$211843 & 10:00:07 & 21:18:43 &  0.21 & FSVS$\_$CL095943$+$210107 & 09:59:43 & 21:01:07 &  0.36 \\
FSVS$\_$CL095906$+$210815 & 09:59:06 & 21:08:15 &  0.38 & FSVS$\_$CL100019$+$210127 & 10:00:19 & 21:01:27 &  0.52 \\
FSVS$\_$CL095857$+$211050 & 09:58:57 & 21:10:50 &  0.54 & FSVS$\_$CL100036$+$215229 & 10:00:36 & 21:52:29 &  0.13 \\
\hline								      
\end{tabular}
\end{table*}

\addtocounter{table}{-1}
\begin{table*}
\center
\caption{Continued.}
\begin{tabular}{llllllll}
\hline
ID  & RA(J2000)  &  Dec(J2000)  &  $z_{est}$  & ID  & RA(J2000)  &  Dec(J2000)  &  $z_{est}$ \\
\hline								      
FSVS$\_$CL100002$+$214502 & 10:00:02 & 21:45:02 &  0.16 & FSVS$\_$CL100042$+$215711 & 10:00:42 & 21:57:11 &  0.16 \\
FSVS$\_$CL100007$+$215028 & 10:00:07 & 21:50:28 &  0.18 & FSVS$\_$CL095853$+$213303 & 09:58:53 & 21:33:03 &  0.36 \\
FSVS$\_$CL095947$+$214953 & 09:59:47 & 21:49:53 &  0.24 & FSVS$\_$CL100045$+$215145 & 10:00:45 & 21:51:45 &  0.27 \\
FSVS$\_$CL100026$+$215049 & 10:00:26 & 21:50:49 &  0.24 & FSVS$\_$CL100039$+$220259 & 10:00:39 & 22:02:59 &  0.32 \\
FSVS$\_$CL100036$+$215038 & 10:00:36 & 21:50:38 &  0.41 & FSVS$\_$CL100041$+$215653 & 10:00:41 & 21:56:53 &  0.32 \\
FSVS$\_$CL100050$+$224858 & 10:00:50 & 22:48:58 &  0.06 & FSVS$\_$CL095852$+$230036 & 09:58:52 & 23:00:36 &  0.19 \\
FSVS$\_$CL095941$+$224640 & 09:59:41 & 22:46:40 &  0.17 & FSVS$\_$CL095922$+$225704 & 09:59:22 & 22:57:04 &  0.16 \\
FSVS$\_$CL100038$+$224949 & 10:00:38 & 22:49:49 &  0.23 & FSVS$\_$CL100007$+$230328 & 10:00:07 & 23:03:28 &  0.31 \\
FSVS$\_$CL095902$+$225638 & 09:59:02 & 22:56:38 &  0.39 & FSVS$\_$CL095943$+$231315 & 09:59:43 & 23:13:15 &  0.37 \\
FSVS$\_$CL100046$+$225357 & 10:00:46 & 22:53:57 &  0.45 & FSVS$\_$CL100103$+$224457 & 10:01:03 & 22:44:57 &  0.46 \\
FSVS$\_$CL100041$+$224931 & 10:00:41 & 22:49:31 &  0.48 & FSVS$\_$CL100015$+$230224 & 10:00:15 & 23:02:24 &  0.48 \\
FSVS$\_$CL100024$+$230153 & 10:00:24 & 23:01:53 &  0.52 & FSVS$\_$CL162423$+$265838 & 16:24:23 & 26:58:38 &  0.12 \\
FSVS$\_$CL162328$+$272231 & 16:23:28 & 27:22:31 &  0.28 & FSVS$\_$CL162423$+$271140 & 16:24:23 & 27:11:40 &  0.34 \\
FSVS$\_$CL162241$+$265902 & 16:22:41 & 26:59:02 &  0.36 & FSVS$\_$CL162417$+$271119 & 16:24:17 & 27:11:19 &  0.37 \\
FSVS$\_$CL162334$+$265947 & 16:23:34 & 26:59:47 &  0.43 & FSVS$\_$CL162430$+$265610 & 16:24:30 & 26:56:10 &  0.57 \\
FSVS$\_$CL162430$+$270645 & 16:24:30 & 27:06:45 &  0.60 & FSVS$\_$CL162308$+$265847 & 16:23:08 & 26:58:47 &  0.79 \\
FSVS$\_$CL162016$+$270018 & 16:20:16 & 27:00:18 &  0.12 & FSVS$\_$CL162115$+$265745 & 16:21:15 & 26:57:45 &  0.18 \\
FSVS$\_$CL162138$+$270643 & 16:21:38 & 27:06:43 &  0.16 & FSVS$\_$CL162020$+$265835 & 16:20:20 & 26:58:35 &  0.20 \\
FSVS$\_$CL162044$+$270340 & 16:20:44 & 27:03:40 &  0.22 & FSVS$\_$CL162030$+$265459 & 16:20:30 & 26:54:59 &  0.25 \\
FSVS$\_$CL162045$+$270739 & 16:20:45 & 27:07:39 &  0.15 & FSVS$\_$CL162031$+$270019 & 16:20:31 & 27:00:19 &  0.26 \\
FSVS$\_$CL162144$+$270952 & 16:21:44 & 27:09:52 &  0.40 & FSVS$\_$CL162012$+$265452 & 16:20:12 & 26:54:52 &  0.34 \\
FSVS$\_$CL162104$+$265842 & 16:21:04 & 26:58:42 &  0.32 & FSVS$\_$CL162038$+$270321 & 16:20:38 & 27:03:21 &  0.56 \\
FSVS$\_$CL162054$+$270003 & 16:20:54 & 27:00:03 &  0.43 & FSVS$\_$CL162057$+$270902 & 16:20:57 & 27:09:02 &  0.72 \\
FSVS$\_$CL162853$+$271110 & 16:28:53 & 27:11:10 &  0.07 & FSVS$\_$CL162851$+$265806 & 16:28:51 & 26:58:06 &  0.11 \\
FSVS$\_$CL162836$+$271322 & 16:28:36 & 27:13:22 &  0.26 & FSVS$\_$CL162901$+$272054 & 16:29:01 & 27:20:54 &  0.37 \\
FSVS$\_$CL162747$+$265225 & 16:27:47 & 26:52:25 &  0.60 & FSVS$\_$CL163130$+$212120 & 16:31:30 & 21:21:20 &  0.14 \\
FSVS$\_$CL163105$+$212141 & 16:31:05 & 21:21:41 &  0.13 & FSVS$\_$CL163110$+$212303 & 16:31:10 & 21:23:03 &  0.14 \\
FSVS$\_$CL163200$+$211445 & 16:32:00 & 21:14:45 &  0.14 & FSVS$\_$CL163141$+$211318 & 16:31:41 & 21:13:18 &  0.15 \\
FSVS$\_$CL163227$+$213308 & 16:32:27 & 21:33:08 &  0.17 & FSVS$\_$CL163147$+$211208 & 16:31:47 & 21:12:08 &  0.19 \\
FSVS$\_$CL163136$+$212955 & 16:31:36 & 21:29:55 &  0.23 & FSVS$\_$CL163217$+$212839 & 16:32:17 & 21:28:39 &  0.21 \\
FSVS$\_$CL163213$+$212342 & 16:32:13 & 21:23:42 &  0.25 & FSVS$\_$CL163156$+$212058 & 16:31:56 & 21:20:58 &  0.33 \\
FSVS$\_$CL163216$+$212541 & 16:32:16 & 21:25:41 &  0.44 & FSVS$\_$CL163215$+$213444 & 16:32:15 & 21:34:44 &  0.58 \\
FSVS$\_$CL071403$+$224833 & 07:14:03 & 22:48:33 &  0.07 & FSVS$\_$CL071535$+$224755 & 07:15:35 & 22:47:55 &  0.20 \\
FSVS$\_$CL071445$+$230251 & 07:14:45 & 23:02:51 &  0.23 & FSVS$\_$CL071403$+$225527 & 07:14:03 & 22:55:27 &  0.26 \\
FSVS$\_$CL071507$+$231123 & 07:15:07 & 23:11:23 &  0.33 & FSVS$\_$CL071401$+$224853 & 07:14:01 & 22:48:53 &  0.50 \\
FSVS$\_$CL071530$+$224527 & 07:15:30 & 22:45:27 &  0.67 & FSVS$\_$CL100055$+$233347 & 10:00:55 & 23:33:47 &  0.08 \\
FSVS$\_$CL100041$+$233440 & 10:00:41 & 23:34:40 &  0.20 & FSVS$\_$CL100044$+$233812 & 10:00:44 & 23:38:12 &  0.36 \\
FSVS$\_$CL100051$+$232624 & 10:00:51 & 23:26:24 &  0.57 & FSVS$\_$CL100055$+$234247 & 10:00:55 & 23:42:47 &  0.57 \\
FSVS$\_$CL100059$+$233521 & 10:00:59 & 23:35:21 &  0.75 & FSVS$\_$CL103624$+$030925 & 10:36:24 & 03:09:25 &  0.13 \\
FSVS$\_$CL103555$+$031910 & 10:35:55 & 03:19:10 &  0.09 & FSVS$\_$CL103635$+$031253 & 10:36:35 & 03:12:53 &  0.13 \\
FSVS$\_$CL103651$+$030422 & 10:36:51 & 03:04:22 &  0.20 & FSVS$\_$CL103644$+$031631 & 10:36:44 & 03:16:31 &  0.21 \\
FSVS$\_$CL103516$+$032348 & 10:35:16 & 03:23:48 &  0.49 & FSVS$\_$CL103858$+$044403 & 10:38:58 & 04:44:03 &  0.08 \\
FSVS$\_$CL103825$+$045409 & 10:38:25 & 04:54:09 &  0.13 & FSVS$\_$CL103834$+$044630 & 10:38:34 & 04:46:30 &  0.19 \\
FSVS$\_$CL103850$+$043145 & 10:38:50 & 04:31:45 &  0.23 & FSVS$\_$CL103821$+$044643 & 10:38:21 & 04:46:43 &  0.26 \\
FSVS$\_$CL103809$+$043323 & 10:38:09 & 04:33:23 &  0.26 & FSVS$\_$CL103842$+$043403 & 10:38:42 & 04:34:03 &  0.26 \\
FSVS$\_$CL103823$+$044346 & 10:38:23 & 04:43:46 &  0.46 & FSVS$\_$CL103756$+$044446 & 10:37:56 & 04:44:46 &  0.56 \\
FSVS$\_$CL103824$+$043750 & 10:38:24 & 04:37:50 &  0.64 & FSVS$\_$CL103813$+$043850 & 10:38:13 & 04:38:50 &  0.79 \\
FSVS$\_$CL172329$+$271820 & 17:23:29 & 27:18:20 &  0.06 & FSVS$\_$CL172242$+$271957 & 17:22:42 & 27:19:57 &  0.10 \\
FSVS$\_$CL172157$+$271708 & 17:21:57 & 27:17:08 &  0.16 & FSVS$\_$CL172158$+$271750 & 17:21:58 & 27:17:50 &  0.15 \\
FSVS$\_$CL172329$+$273412 & 17:23:29 & 27:34:12 &  0.19 & FSVS$\_$CL172318$+$272032 & 17:23:18 & 27:20:32 &  0.24 \\
FSVS$\_$CL172318$+$272018 & 17:23:18 & 27:20:18 &  0.24 & FSVS$\_$CL172159$+$271857 & 17:21:59 & 27:18:57 &  0.40 \\
FSVS$\_$CL172335$+$274046 & 17:23:35 & 27:40:46 &  0.40 & FSVS$\_$CL172235$+$272616 & 17:22:35 & 27:26:16 &  0.43 \\
FSVS$\_$CL172258$+$273139 & 17:22:58 & 27:31:39 &  0.68 & FSVS$\_$CL172249$+$273207 & 17:22:49 & 27:32:07 &  0.66 \\
FSVS$\_$CL172326$+$272459 & 17:23:26 & 27:24:59 &  0.72 & FSVS$\_$CL172253$+$272425 & 17:22:53 & 27:24:25 &  0.87 \\
FSVS$\_$CL172610$+$271749 & 17:26:10 & 27:17:49 &  0.07 & FSVS$\_$CL172607$+$271544 & 17:26:07 & 27:15:44 &  0.12 \\
FSVS$\_$CL172431$+$272208 & 17:24:31 & 27:22:08 &  0.13 & FSVS$\_$CL172623$+$272637 & 17:26:23 & 27:26:37 &  0.10 \\
FSVS$\_$CL172540$+$273314 & 17:25:40 & 27:33:14 &  0.15 & FSVS$\_$CL172431$+$271507 & 17:24:31 & 27:15:07 &  0.20 \\
FSVS$\_$CL172425$+$272530 & 17:24:25 & 27:25:30 &  0.35 & FSVS$\_$CL172427$+$273158 & 17:24:27 & 27:31:58 &  0.29 \\
FSVS$\_$CL172416$+$272230 & 17:24:16 & 27:22:30 &  0.28 & FSVS$\_$CL172624$+$272559 & 17:26:24 & 27:25:59 &  0.42 \\
FSVS$\_$CL172531$+$274247 & 17:25:31 & 27:42:47 &  0.60 & FSVS$\_$CL172427$+$272306 & 17:24:27 & 27:23:06 &  0.59 \\
\hline
\end{tabular}
\end{table*}

\addtocounter{table}{-1}
\begin{table*}
\center
\caption{Continued.}
\begin{tabular}{llllllll}
\hline
ID  & RA(J2000)  &  Dec(J2000)  &  $z_{est}$  & ID  & RA(J2000)  &  Dec(J2000)  &  $z_{est}$ \\
\hline								      
FSVS$\_$CL172556$+$274145 & 17:25:56 & 27:41:45 &  0.82 & FSVS$\_$CL172709$+$271615 & 17:27:09 & 27:16:15 &  0.13 \\
FSVS$\_$CL172742$+$273139 & 17:27:42 & 27:31:39 &  0.47 & FSVS$\_$CL172742$+$273144 & 17:27:42 & 27:31:44 &  0.60 \\
FSVS$\_$CL172824$+$273109 & 17:28:24 & 27:31:09 &  0.65 & FSVS$\_$CL215855$+$272406 & 21:58:55 & 27:24:06 &  0.13 \\
FSVS$\_$CL215947$+$275205 & 21:59:47 & 27:52:05 &  0.16 & FSVS$\_$CL215937$+$272756 & 21:59:37 & 27:27:56 &  0.50 \\
FSVS$\_$CL220015$+$273208 & 22:00:15 & 27:32:08 &  0.77 & FSVS$\_$CL220214$+$275310 & 22:02:14 & 27:53:10 &  0.25 \\
FSVS$\_$CL220301$+$272628 & 22:03:01 & 27:26:28 &  0.41 & FSVS$\_$CL220246$+$272756 & 22:02:46 & 27:27:56 &  0.71 \\
FSVS$\_$CL220443$+$274627 & 22:04:43 & 27:46:27 &  0.21 & FSVS$\_$CL220503$+$274751 & 22:05:03 & 27:47:51 &  0.22 \\
FSVS$\_$CL220530$+$273041 & 22:05:30 & 27:30:41 &  0.27 & FSVS$\_$CL220537$+$272418 & 22:05:37 & 27:24:18 &  0.29 \\
FSVS$\_$CL220512$+$272614 & 22:05:12 & 27:26:14 &  0.36 & FSVS$\_$CL220537$+$272727 & 22:05:37 & 27:27:27 &  0.34 \\
FSVS$\_$CL220424$+$272631 & 22:04:24 & 27:26:31 &  0.51 & FSVS$\_$CL220427$+$272551 & 22:04:27 & 27:25:51 &  0.51 \\
FSVS$\_$CL220421$+$272605 & 22:04:21 & 27:26:05 &  0.71 & FSVS$\_$CL220431$+$271124 & 22:04:31 & 27:11:24 &  0.16 \\
FSVS$\_$CL220506$+$271052 & 22:05:06 & 27:10:52 &  0.33 & FSVS$\_$CL220431$+$270704 & 22:04:31 & 27:07:04 &  0.51 \\
FSVS$\_$CL220456$+$271659 & 22:04:56 & 27:16:59 &  0.46 & FSVS$\_$CL220532$+$270715 & 22:05:32 & 27:07:15 &  0.49 \\
FSVS$\_$CL220436$+$270653 & 22:04:36 & 27:06:53 &  0.57 & FSVS$\_$CL220422$+$265841 & 22:04:22 & 26:58:41 &  0.68 \\
FSVS$\_$CL182829$+$355431 & 18:28:29 & 35:54:31 &  0.16 & FSVS$\_$CL182900$+$360121 & 18:29:00 & 36:01:21 &  0.13 \\
FSVS$\_$CL183017$+$361045 & 18:30:17 & 36:10:45 &  0.26 & FSVS$\_$CL182817$+$355616 & 18:28:17 & 35:56:16 &  0.30 \\
FSVS$\_$CL182919$+$360440 & 18:29:19 & 36:04:40 &  0.21 & FSVS$\_$CL183020$+$361045 & 18:30:20 & 36:10:45 &  0.30 \\
FSVS$\_$CL182951$+$360210 & 18:29:51 & 36:02:10 &  0.28 & FSVS$\_$CL182921$+$355004 & 18:29:21 & 35:50:04 &  0.34 \\
FSVS$\_$CL182835$+$355108 & 18:28:35 & 35:51:08 &  0.32 & FSVS$\_$CL182949$+$354719 & 18:29:49 & 35:47:19 &  0.34 \\
FSVS$\_$CL182906$+$354708 & 18:29:06 & 35:47:08 &  0.43 & FSVS$\_$CL182837$+$360204 & 18:28:37 & 36:02:04 &  0.38 \\
FSVS$\_$CL182816$+$355813 & 18:28:16 & 35:58:13 &  0.48 & FSVS$\_$CL183019$+$360124 & 18:30:19 & 36:01:24 &  0.43 \\
FSVS$\_$CL183004$+$355143 & 18:30:04 & 35:51:43 &  0.75 & FSVS$\_$CL183035$+$355717 & 18:30:35 & 35:57:17 &  0.76 \\
FSVS$\_$CL183042$+$355456 & 18:30:42 & 35:54:56 &  0.83 & FSVS$\_$CL183126$+$354858 & 18:31:26 & 35:48:58 &  0.16 \\
FSVS$\_$CL183228$+$355929 & 18:32:28 & 35:59:29 &  0.20 & FSVS$\_$CL183220$+$360823 & 18:32:20 & 36:08:23 &  0.15 \\
FSVS$\_$CL183310$+$355802 & 18:33:10 & 35:58:02 &  0.20 & FSVS$\_$CL183223$+$360901 & 18:32:23 & 36:09:01 &  0.18 \\
FSVS$\_$CL183158$+$354612 & 18:31:58 & 35:46:12 &  0.33 & FSVS$\_$CL183214$+$361227 & 18:32:14 & 36:12:27 &  0.47 \\
FSVS$\_$CL183121$+$354556 & 18:31:21 & 35:45:56 &  0.63 & FSVS$\_$CL183342$+$355710 & 18:33:42 & 35:57:10 &  0.71 \\
FSVS$\_$CL183326$+$354927 & 18:33:26 & 35:49:27 &  0.81 & FSVS$\_$CL183258$+$360953 & 18:32:58 & 36:09:53 &  0.76 \\
FSVS$\_$CL183414$+$364833 & 18:34:14 & 36:48:33 &  0.21 & FSVS$\_$CL183413$+$363138 & 18:34:13 & 36:31:38 &  0.15 \\
FSVS$\_$CL183414$+$363132 & 18:34:14 & 36:31:32 &  0.12 & FSVS$\_$CL183341$+$363447 & 18:33:41 & 36:34:47 &  0.28 \\
FSVS$\_$CL183404$+$364722 & 18:34:04 & 36:47:22 &  0.22 & FSVS$\_$CL183319$+$364852 & 18:33:19 & 36:48:52 &  0.24 \\
FSVS$\_$CL183356$+$364625 & 18:33:56 & 36:46:25 &  0.32 & FSVS$\_$CL183417$+$364924 & 18:34:17 & 36:49:24 &  0.34 \\
FSVS$\_$CL183331$+$364908 & 18:33:31 & 36:49:08 &  0.35 & FSVS$\_$CL183311$+$364042 & 18:33:11 & 36:40:42 &  0.41 \\
FSVS$\_$CL183316$+$362402 & 18:33:16 & 36:24:02 &  0.43 & FSVS$\_$CL183331$+$364906 & 18:33:31 & 36:49:06 &  0.35 \\
FSVS$\_$CL183407$+$363917 & 18:34:07 & 36:39:17 &  0.50 & FSVS$\_$CL183316$+$363622 & 18:33:16 & 36:36:22 &  0.41 \\
FSVS$\_$CL183349$+$363729 & 18:33:49 & 36:37:29 &  0.43 & FSVS$\_$CL183320$+$363649 & 18:33:20 & 36:36:49 &  0.44 \\
FSVS$\_$CL183252$+$363041 & 18:32:52 & 36:30:41 &  0.47 & FSVS$\_$CL183406$+$363919 & 18:34:06 & 36:39:19 &  0.52 \\
FSVS$\_$CL183328$+$363045 & 18:33:28 & 36:30:45 &  0.50 & FSVS$\_$CL183321$+$363357 & 18:33:21 & 36:33:57 &  0.70 \\
FSVS$\_$CL183318$+$353805 & 18:33:18 & 35:38:05 &  0.13 & FSVS$\_$CL183158$+$352307 & 18:31:58 & 35:23:07 &  0.11 \\
FSVS$\_$CL183219$+$352308 & 18:32:19 & 35:23:08 &  0.14 & FSVS$\_$CL183115$+$351504 & 18:31:15 & 35:15:04 &  0.22 \\
FSVS$\_$CL183246$+$353829 & 18:32:46 & 35:38:29 &  0.32 & FSVS$\_$CL183208$+$352628 & 18:32:08 & 35:26:28 &  0.16 \\
FSVS$\_$CL183211$+$351338 & 18:32:11 & 35:13:38 &  0.30 & FSVS$\_$CL183204$+$354040 & 18:32:04 & 35:40:40 &  0.43 \\
FSVS$\_$CL183247$+$353902 & 18:32:47 & 35:39:02 &  0.46 & FSVS$\_$CL183221$+$352625 & 18:32:21 & 35:26:25 &  0.59 \\
FSVS$\_$CL183234$+$352605 & 18:32:34 & 35:26:05 &  0.53 & FSVS$\_$CL183300$+$351934 & 18:33:00 & 35:19:34 &  0.56 \\
FSVS$\_$CL183318$+$353950 & 18:33:18 & 35:39:50 &  0.69 & FSVS$\_$CL183122$+$351357 & 18:31:22 & 35:13:57 &  0.76 \\
FSVS$\_$CL183215$+$352818 & 18:32:15 & 35:28:18 &  0.81 & FSVS$\_$CL234225$+$280950 & 23:42:25 & 28:09:50 &  0.08 \\
FSVS$\_$CL234052$+$275951 & 23:40:52 & 27:59:51 &  0.20 & FSVS$\_$CL234308$+$280238 & 23:43:08 & 28:02:38 &  0.14 \\
FSVS$\_$CL234257$+$280904 & 23:42:57 & 28:09:04 &  0.22 & FSVS$\_$CL234208$+$281321 & 23:42:08 & 28:13:21 &  0.17 \\
FSVS$\_$CL234214$+$275634 & 23:42:14 & 27:56:34 &  0.50 & FSVS$\_$CL234156$+$282029 & 23:41:56 & 28:20:29 &  0.58 \\
FSVS$\_$CL234357$+$280649 & 23:43:57 & 28:06:49 &  0.10 & FSVS$\_$CL234353$+$280421 & 23:43:53 & 28:04:21 &  0.14 \\
FSVS$\_$CL234424$+$280709 & 23:44:24 & 28:07:09 &  0.22 & FSVS$\_$CL234416$+$281315 & 23:44:16 & 28:13:15 &  0.21 \\
FSVS$\_$CL234358$+$281138 & 23:43:58 & 28:11:38 &  0.26 & FSVS$\_$CL234457$+$281547 & 23:44:57 & 28:15:47 &  0.26 \\
FSVS$\_$CL234515$+$280939 & 23:45:15 & 28:09:39 &  0.27 & FSVS$\_$CL234450$+$281931 & 23:44:50 & 28:19:31 &  0.57 \\
FSVS$\_$CL234506$+$282446 & 23:45:06 & 28:24:46 &  0.57 & FSVS$\_$CL234458$+$281611 & 23:44:58 & 28:16:11 &  0.68 \\
FSVS$\_$CL234806$+$281205 & 23:48:06 & 28:12:05 &  0.20 & FSVS$\_$CL234821$+$281741 & 23:48:21 & 28:17:41 &  0.15 \\
FSVS$\_$CL234656$+$275925 & 23:46:56 & 27:59:25 &  0.27 & FSVS$\_$CL234824$+$281851 & 23:48:24 & 28:18:51 &  0.34 \\
FSVS$\_$CL234740$+$275716 & 23:47:40 & 27:57:16 &  0.42 & FSVS$\_$CL234754$+$280913 & 23:47:54 & 28:09:13 &  0.41 \\
FSVS$\_$CL234806$+$281758 & 23:48:06 & 28:17:58 &  0.46 & FSVS$\_$CL234807$+$281734 & 23:48:07 & 28:17:34 &  0.51 \\
FSVS$\_$CL234808$+$281135 & 23:48:08 & 28:11:35 &  0.55 & FSVS$\_$CL234745$+$275720 & 23:47:45 & 27:57:20 &  0.49 \\
FSVS$\_$CL234632$+$281259 & 23:46:32 & 28:12:59 &  0.58 & FSVS$\_$CL234904$+$281231 & 23:49:04 & 28:12:31 &  0.09 \\
\hline
\end{tabular}
\end{table*}

\addtocounter{table}{-1}
\begin{table*}
\center
\caption{Continued.}
\begin{tabular}{llllllll}
\hline
ID  & RA(J2000)  &  Dec(J2000)  &  $z_{est}$  & ID  & RA(J2000)  &  Dec(J2000)  &  $z_{est}$ \\
\hline								      
FSVS$\_$CL234953$+$282735 & 23:49:53 & 28:27:35 &  0.15 & FSVS$\_$CL235000$+$283401 & 23:50:00 & 28:34:01 &  0.16 \\
FSVS$\_$CL234916$+$281246 & 23:49:16 & 28:12:46 &  0.21 & FSVS$\_$CL235019$+$281756 & 23:50:19 & 28:17:56 &  0.29 \\
FSVS$\_$CL235052$+$281135 & 23:50:52 & 28:11:35 &  0.22 & FSVS$\_$CL235011$+$283120 & 23:50:11 & 28:31:20 &  0.28 \\
FSVS$\_$CL235104$+$281057 & 23:51:04 & 28:10:57 &  0.33 & FSVS$\_$CL235024$+$281500 & 23:50:24 & 28:15:00 &  0.46 \\
FSVS$\_$CL235013$+$280723 & 23:50:13 & 28:07:23 &  0.35 & FSVS$\_$CL235012$+$280716 & 23:50:12 & 28:07:16 &  0.41 \\
FSVS$\_$CL235102$+$281304 & 23:51:02 & 28:13:04 &  0.47 & FSVS$\_$CL235031$+$282856 & 23:50:31 & 28:28:56 &  0.50 \\
\hline
\end{tabular}
\end{table*}

The FSVS cluster catalogue can be accessed at\\
http://www-astro.physics.ox.ac.uk/$\sim$iks/FSVScatalogue/ home.html with a
further distribution planned to be through the NOAO data products pages at
http://www.noao.edu/ dpp/ . The on-line version of the FSVS cluster catalogue
contains the full information in the following columns:
(1)  number of the FSVS field in which the cluster has been detected;
(2)  identification, ID, of the cluster, formed from the survey name
     (FSVS$\_$CL) and the coordinates;
(3)  Right Ascension, RA, equinox J2000, of the cluster centre, defined as
     the mean position of its members; 
(4)  Declination, Dec, equinox J2000, of the cluster centre, defined as
     the mean position of its members; 
(5)  number of member galaxies, $N_{m}$, defined as the number of all objects
     found within the boundary of the cluster and falling within the
     colour-magnitude filter in which the cluster has been detected;
(6)  $N_{all}$, number of all objects contained within the cluster boundary;
(7)  $A_{cl}$, area occupied by the cluster in arcmin$^{2}$;
(8)  projected number density of member galaxies in arcmin$^{-2}$;
(9)  BCG~$I$ magnitude of the brightest cluster galaxy;
(10) $m_{3}$, $I$ magnitude of the third brightest member galaxy;
(11) $C_{fl}$, maximum colour of the colour-magnitude filter in which the
     cluster has been detected;
(12) richness, $R_{m}$, of the cluster, defined as the number of galaxies found
     within its spatial boundary, its colour filter, and the magnitude
     range $m_{3} < I < m_{3}+2$, where $m_{3}$ is the magnitude of the third
     brightest member;
(13) flag indicating if the richness is underestimated because of a restricted
     magnitude range;
(14) estimated redshift, $z_{est}$;
(15) flag indicating if the parameters of the cluster may have been influenced
     by the field boundary;
(16) zero-point of the CRS normalised to $I=17$ mag;
(17) slope of the CRS.

In addition to the tabular data, a set of complementary files can be accessed
for every cluster. These files include: (1) an ASCII table of all objects
contained within the cluster boundary, with a flag for objects which are
cluster members, plus the coordinates of the cluster boundary points; (2) a
contour plot of the galaxy number density in the cluster field with
delineated boundary of the cluster and marked positions of the member
galaxies; (3) a colour-magnitude diagram of objects within the cluster
boundary and cluster members, plus the fit of the CRS.

\section{Summary}

Clusters of galaxies are a powerful tool in the study of cosmological
models. However, large uncertainties arise from our still limited
understanding of cluster formation and evolution. The availability of
appropriate cluster samples is the basic requirement for systematic and
meaningful studies of the various processes involved in cluster formation and
in their evolution with redshift. The properties of such samples would be:
(1) a wide range of redshifts to resolve the evolutionary changes; (2) a wide
range of masses in terms of both the number of member galaxies and gas
content; (3) a variety of cluster morphologies; and (4) good accessibilty for
follow-up observations. Fulfilling all of the above requirements, the FSVS
cluster sample is exceptionally well suited to studying the formation and
evolution of galaxy clusters.

The 598 clusters presented here have been identified using a fully automated,
semi-parametric technique based on a maximum likelihood approach applied to
Voronoi tessellation, and enhanced by colour discrimination. It is a
morphologically-unbiased sample, containing structures with a wide range of
richnesses and evolutionary stages, with a mean density of $\sim 28$ clusters
deg$^{-2}$, and spanning a range of estimated redshifts of $0.05 < z < 0.9$
with mean $\langle z \rangle = 0.345$. The contamination of the catalogue by
spurious detections is estimated to be $< 24 \%$ and restricted mainly to the
subsample of the low redshift ($z_{est} < 0.4$) groups and very poor
clusters.

The redshifts are estimated assuming the homogeneity of the CRS for clusters
at the same redshift. The redshift uncertainty of the estimated redshifts is
$\sigma = 0.03$. The comparison with the galaxy clusters from Stanford et
al.\ (2002) indicates that the estimated redshifts from the CRS could be
slightly biased (an offset of $+0.014$). However, this offset could instead
arise from the lower-quality photometry available for the Stanford clusters.

The catalogue here (Table~\ref{catalogue}) contains only the most basic
parameters for every cluster and is intended simply as a quick reference. The
full parameter list and all associated data will be released on-line through
NOAO data products. The users of the FSVS cluster catalogue are advised to
consult the on-line pages for any subsequent improvements (e.g.\ more
accurate redshifts), changes, and data additions.

\section*{Acknowledgments}

IKS was supported by a Marie Curie Fellowship, Improving Human Potential,
contract HPMD-CT-2000-00005. We would like to acknowledge the use of the NED
and Vizier facilities. We would like to thank the FSVS consortium for
preparation of the data product. The FSVS is part of the INT Wide Field
Survey. The INT and WHT are operated on the island of La Palma by the Isaac
Newton Group in the Spanish Observatorio del Roque de los Muchachos of the
Inst\'{\i}tuto de Astrof\'{\i}sica de Canarias. We would like to particularly
acknowledge the Service programme at the WHT, which allowed us to acquire
spectroscopic redshifts in a timely manner. We would like to thank the
anonymous referee for helpful comments.

This work was in part performed under the auspices of the US Department of
Energy, National Nuclear Security Administration via the University of
California, Lawrence Livermore National Laboratory, under contract no.\
W-7405-End-48.

\bsp

\label{lastpage}


\begin{thebibliography}{99}

\bibitem [\protect\citename{Abell }1958]{Ab58}
Abell G. O., 1958, ApJS, 3, 211
%
\bibitem[\protect\citename{Allard \& Fraley }1997]{AF97}
Allard D., Fraley C., 1997, JASA, 92, 1485
%
\bibitem [\protect\citename{Bertin \& Arnouts }1996]{BA96}
Bertin E., Arnouts S., 1996, A\&AS, 117, 393
%
\bibitem [\protect\citename{Colless et al.\ }2001]{Co01}
Colless M., et al., 2001, MNRAS, 328, 1039
%
\bibitem [\protect\citename{Dodd \& MacGillivray }1986]{DM86}
Dodd R. J., MacGillivray H. T., 1986, AJ, 92, 706
%
\bibitem [\protect\citename {Gal et al.\ }2000] {Ga00}
Gal R. R., de Carvalho R. R., Odewahn S. C., Djorgovski S. G.,
Margoniner V. E., 2000, AJ, 119, 12
%
\bibitem [\protect\citename{Gladders \& Yee }2000]{GY00}
Gladders M., Yee H. K. C., 2000, AJ, 120, 2148
%
\bibitem [\protect\citename{Goto et al.\ }2002]{Go02}
Goto T., et al., 2002, AJ, 123, 1807
%
\bibitem [\protect\citename{Groot et al.\ }2003]{Gr03}
Groot P. J., et al., 2003, MNRAS, 339, 427 
%
\bibitem [\protect\citename{Hansen et al.\ }2005]{ha05}
Hansen, Sarah M.; McKay, Timothy A.; Wechsler, Risa H.; Annis, James; Sheldon, Erin Scott; Kimball, Amy, 2005, ApJ, 633, 122
%
\bibitem [\protect\citename{Haines et al.\ }2001]{ha}
Haines C. P., Clowes R. G., Campusano L. E., Adamson A. J., 2001,
MNRAS, 323, 688
%
\bibitem [\protect\citename{Hawkins et al.\ }2003]{Ha03}
Hawkins E., et al.\, 2003, MNRAS, 346, 78
%
\bibitem [\protect\citename{Huber }2002]{Hu02}
Huber M., 2002, PhD Thesis, University of Wyoming
%
\bibitem [\protect\citename{Jones et al.\ }1991]{Jo91}
Jones L. R., Fong R., Shanks T., Ellis R. S., Peterson B. A., 1991,
MNRAS, 249, 481
%
\bibitem [\protect\citename {Kepner et al.\ }1999]{Ke99}
Kepner J., Fan X., Bahcall N., Gunn J., Lupton R., Xu G., 1999, ApJ, 517, 78
%
\bibitem[\protect\citename{Kim et al.\ }2000]{Ki00}
Kim R. S. J., et al., 2000, in Mazure, Le F\`evre, Le Brun,
ASP Conference Series, Vol. 200, p. 422
%
\bibitem[\protect\citename{Kim et al.\ }2002]{Ki02}
Kim R. S. J., et al., 2002, AJ, 123, 20
%
\bibitem [\protect\citename {Kodama et al.\ }1998]{Ko98}
Kodama T., Arimoto N., Barger A. J., Arag\'{o}n-Salamanca A., 1998,
A\&A, 334, 99
%
\bibitem [\protect\citename {Lin \& Mohr }2004]{Li04}
Lin Y., Mohr J.J., 2005, ApJ, 617, 879
%
\bibitem [\protect\citename {Mullis et al.\ }2003]{Mu03} Mullis C.R.,
McNamara B.R., Quintana H., Vikhlinin A., Henry J.P., Gioia I.M., Hornstrup
A., Forman W., Jones C., 2003, ApJ, 594, 154
%
\bibitem [\protect\citename {Novicki et al.\ }2002]{No02}
Novicki M.C., Sornig M., Henry J.P., 2002, AJ, 124, 2413
%
\bibitem [\protect\citename{Okabe et al.\ }2000]{ok}
Okabe A., Boots B., Sugihara K., Chiu S. N., 2000, Spatial Tessellations,
2nd edn. John Wiley \& Sons
%
\bibitem[\protect\citename{Parker et al.\ }2003]{Pa03}
Parker L., Komp W., Vanzella D. A. T., 2003, ApJ, 588, 663
%
\bibitem[\protect\citename{Percival et al.\ }2002]{Pe02}
Percival W. J., et al.\, 2002, MNRAS, 337, 1068
%
\bibitem[\protect\citename{Popesso et al.\ }2004]{Po04} Popesso P.,
Boehringer H., Brinkmann J., Voges W., York D.G., 2004, A\&A, 423, 449
%
\bibitem [\protect\citename{Postman et al.\ }1996]{Po96}
Postman M., Lubin L., Gunn J. E., Oke J. B., Hoessel J. G., Schneider D. P.,
Christensen J. A., 1996, AJ, 111, 615
%
\bibitem [\protect\citename{Press \& Schechter }1974]{PS74}
Press W. H., Schechter P., 1974, ApJ, 187, 425
%
\bibitem [\protect\citename{Ramella et al.\ }2001]{Ra01}
Ramella M., Boschin W., Fadda D., Nonino M., 2001, A\&A, 308, 776
%
\bibitem [\protect\citename{S\'{a}nchez \& Gomz\'{a}les-Serrano }2002]{SG02}
S\'{a}nchez S. F., Gomz\'{a}les-Serrano J. I., 2002, astro-ph/0209355
%
\bibitem [\protect\citename{Shectman }1985]{Sh85}
Shectman S. A., 1985, ApJS, 57, 77
%
\bibitem[\protect\citename{S\"{o}chting }2002]{So02a}
S\"{o}chting I. K., 2002, PhD Thesis, University of Central Lancashire
%
\bibitem[\protect\citename{S\"{o}chting, Clowes \& Campusano }2002]{So02}
S\"{o}chting I. K., Clowes R. G., Campusano L. E., 2002, MNRAS, 331, 569
%
\bibitem[\protect\citename{S\"{o}chting, Clowes \& Campusano }2003]{So03}
S\"{o}chting I. K., Clowes R. G., Campusano L. E., 2004, MNRAS, 347, 1241
%
\bibitem[\protect\citename{Spergel et al.\ }2003]{Sp03}
Spergel D. N. et al., 2003, ApJS, 148, 175
%
\bibitem[\protect\citename{Stanford et al.\ }2002]{St02} Stanford S. A.,
Eisenhardt P. R., Dickinson M., Holden B. P., De Propris R., 2002, ApJS, 142,
153
%
\bibitem[\protect\citename{Valotto, Moore \& Lambas }2001]{Va01}
Valotto C. A., Moore B., Lambas D. G., 2001, ApJ, 546, 157
%
\bibitem[\protect\citename{Viana \& Liddle }1999]{VL99}
Viana P. T., Liddle A. R., 1999, MNRAS, 303, 535
%
\bibitem [\protect\citename{Zwicky et al.\ }1961--1968]{Zw61}
Zwicky F., Herzog E., Wild P., Karpowicz M., Kowal C. T., 1961--1968, Catalog
of Galaxies and Clusters of Galaxies, in 6 vols., Pasadena, California
Institute of Technology

\end{thebibliography}
\end{document}